\documentclass[]{aa}

\usepackage{graphicx}
\usepackage[colorlinks,allcolors=blue]{hyperref}
\usepackage{txfonts}
\newcommand{\kms}{km\,s$^{-1}$}

\begin{document}

   \title{Star formation quenching stages of active and non-active galaxies}
   \subtitle{}

   \author{V.  Kalinova\inst{1}\fnmsep\thanks{email: kalinova@mpifr.de}
          \and D. Colombo\inst{1}
          \and S. F. S{\'a}nchez\inst{2}
          \and K. Kodaira\inst{1,3,4}
          \and R. Garc{\'i}a-Benito\inst{5}
          \and R. Gonz{\'a}lez Delgado\inst{5}
          \and E. Rosolowsky\inst{6}
          \and E. A. D. Lacerda\inst{2}
          }

   \institute{Max Planck Institute for Radioastronomy, Auf dem H\"ugel 69, D-53121 Bonn, Germany
        \and Instituto de Astronom\'\i a,Universidad Nacional Auton\'oma de M\'exico,   A.P. 70-264, 04510, M\'exico, D.F.
         \and National Astronomical Observatory of Japan;  Osawa2-21-1, Mitaka-shi, Tokyo, Japan  PC 181-8588
         \and SOKENDAI, International Village, Hayama-machi, Miura-gun, Kanagawa-ken, Japan, PC 240-0193
         \and Instituto de Astrof\'isica de Andaluc\'ia, CSIC, Apartado de correos 3004, 18080 Granada, Spain 
         \and Department of Physics 4-181 CCIS, University of Alberta, Edmonton AB T6G 2E1, Canada 
             }

   \date{Received November 13, 2020; accepted January 21, 2021}

 
  \abstract{
The mechanisms that bring galaxies to strongly reduce their star formation activity (star-formation quenching) is still poorly understood. To better study galaxy evolution, we propose a classification based on the maps of the ionised hydrogen distribution, traced by kpc-resolved, equivalent width of H$\alpha$ maps, and the nuclear activity of the galaxies using information from the Baldwin-Philips-Terlevich (BPT) diagnostic diagrams. Using these tools, we group a sample of 238 galaxies from the CALIFA (Calar Alto Legacy Integral Field spectroscopy Area) survey in six quenching stages (QS):  (i) objects dominated by recent star formation; (ii) systems that present a quiescent-nuclear-ring structure in their centre; (iii) galaxies that are centrally-quiescent;  (iv) galaxies with no clear pattern in their ionisation gas distribution $-$ mixed; (v) systems that posses only a few star-forming regions $-$ nearly-retired, or (vi) galaxies that are completely quiescent $-$ fully-retired.  Regarding their nuclear activity, we further divide the galaxies into two groups $-$ active systems that host a weak or strong Active Galactic Nuclei (AGN) in their centre, and non-active objects.
Galaxies grouped into quenching stage classes occupy specific locations on the star-formation-rate versus stellar mass diagram.
The ``Blue cloud'' is populated by the star-forming and the quiescent-nuclear-ring galaxies, the ``Green valley'' is populated by centrally-quiescent and mixed systems, ``Red sequence'' by the nearly- and fully-retired objects.
Generally, galaxies that host a weak or strong AGN show properties, comparable to the non-active counterparts at the same quenching stages, except for the AGN-hosting star-forming systems. The degree of the star-formation quenching increases along the present emission-line pattern sequence from star-forming to fully-retired.
The proposed emission-line classes reinforce the "inside-out" quenching scenario, which foresees that the suppression of the star-formation begins from the central regions of the galaxies. }


   \maketitle
%


\defcitealias{Kalinova2017b}{K17}


\section{Introduction}
\label{S:intro}
Distributions of galaxy properties such as morphologies (\citealt{Hubble1926}), colours (\citealt{Strateva2001}, \citealt{Baldry2004}), star-formation rates (\citealt{Brinchmann2004}, \citealt{Renzini2015}), ages (\citealt{Gallazzi2005,Gallazzi2008,Zibetti2017}) and gas content (e.g., \citealt{Young991}, \citealt{Blanton2009}) in the local universe show strong bi-modality. Similar bi-modality is observed even up to $z\sim$ 2.5 (\citealt{Brammer2009, Williams2009}) . The so-called ``red sequence'' is populated by red spheroidal systems, while the ``blue cloud'' hosts blue disk galaxies. The region in between is called the ``green valley'' and it is underpopulated, suggesting that the transition occurs within a narrow range of certain galaxy parameters from the blue cloud to the red sequence. Galaxies in the green valley do not show a peculiar morphology (\citealt{Schawinski2014}) as well as other particular properties, rather their colour is the consequence of the switching off of the star formation, a process usually called ``star formation quenching'' 
\citep[e.g., ][]{Faber2007}. Indeed, when posed on a star formation rate (SFR) versus stellar mass ($M_*$) diagram, galaxies show the same bi-modality, with star-forming galaxies tightly organised across the ``star formation main sequence'' \citep[SFMS, e.g. ][]{Brinchmann2004, Whitaker2012, Renzini2015, Speagle2014, Cano-Diaz2016,Sanchez2019}, retired galaxies in the corresponding ``red sequence'', and galaxies in the process of quenching (immediately below the main sequence) in the ``green valley'' (e.g., \citealt{Schawinski2014}).  

The integrated star formation rate versus stellar mass diagram (SFR-M$_*$) has been broadly explored in the literature 
(e.g., \citealt{Schawinski2014}; \citealt{Renzini2015}; \citealt{Catalan-Torrecilla2015}; \citealt{Gonzalez-Delgado2016}; \citealt{Cano-Diaz2016}; \citealt{Catalan-Torrecilla2017}; \citealt{Belfiore2018}; \citealt{Cano-Diaz2019}; \citealt{Bluck2019}; \citealt{Sanchez2018,Sanchez2019}; \citealt{Lacerda2020}) and has served as an important diagnostic tool for studying of galaxy evolution.

Additionally, galaxy evolution theories have been largely constrained through ionised gas emission-line (EL) classification schemes that study star formation, chemical properties and nuclear activity of large samples of galaxies. \cite{Baldwin1981} \citep[see also ][]{Veilleux1987} originally proposed a diagram (now known as the BPT diagram) that uses the intensity ratio of emission lines ([O~III] $\lambda$5007/H$\beta$ versus [N~II] $\lambda$6584/H$\alpha$ line ratios) to distinguish between dominant excitation sources (such as HII regions, power-law continuum spectrum photo-ionisations, and heating by shock-waves in the original formulation) from the spectra of extragalactic objects. Later on \cite{Kauffmann2003} defined loci on the BPT diagram to distinguish  star-forming and star-burst galaxies \citep[see also ][]{Kewley2001}, Seyfert regions related to galaxies hosting an Active Galactic Nuclei (AGN), or low ionisation nuclear emission regions (LINERs), in which the ionisation might be attributed to old and hot stars (e.g., \citealt{Stasinska2008}).

More recently, \cite{CidFernandes2011} proposed a different bi-dimensional diagram to disentangle the excitation source in the centre of galaxies. The ``WHAN'' diagram consider the equivalent width of the H$\alpha$ versus the [N~II] $\lambda$6584/H$\alpha$ ratio to classify galaxies into purely star-forming or strong AGN-hosts. Additionally, this diagnostic is able to distinguish between objects that are retired (or passive) from galaxies that are genuinely hosting a weak AGN.    

The advent of spatially-resolved, Integral Field Spectroscopy (IFS) surveys such as ATLAS$^{\rm 3D}$ (\citealt{cappellari11}),
CALIFA (Calar Alto Legacy Integral Field spectroscopy Area; \citealt{Sanchez2012}), SAMI (Sydney-AAO Multi-object Integral field spectrograph; \citealt{Croom2012}), MaNGA (Mapping Nearby Galaxies at Apache Point Observatory; \citealt{Bundy2015}), AMUSING++ (All-weather MUse Supernova Integral field Nearby Galaxies; \citealt{Lopez-Coba2020}) has provided large, statistical samples to study excitation sources of the ionised gas not only between galaxies, but also within them. In particular, they give the opportunity to identify the dominant excitation source of a galaxy, since different galactic regions can be characterised by a large variety of ionisation effects (see \citealt{Sanchez2020} for an extensive review).

An attempt to classify single regions of galaxies through the use of the BPT diagram has been performed by \cite{Singh2013}, who show that excitation properties that characterise LINERs are not simply limited to the nuclear regions, but they extend far away from galactic centre. Those ``low ionisation emission regions'' (therefore redefined as ``LIERs'') are probably due to post-AGB stars, rather than to AGN activity.
Following a similar method, \cite{Belfiore2016} used the BPT scheme to group  galaxies based on their dominant excitation source and consider the spatially resolved patterns described by the diagram. This classification distinguishes between galaxies dominated by star formation, galaxies host of an AGN, objects that show LIERs only in the centre (cLIERs) or throughout their full extent (eLIERs), as well as mergers. In particular cLIERs are bulged spiral galaxies located in the green valley, while eLIERs are basically elliptical, retired objects. 

The diffuse ionised gas classification proposed by \cite{Lacerda2018}, instead, put emphasis on the use of resolved H$\alpha$ equivalent width maps to show where the gas excitation in the different galaxy regions is powered by hot low-mass evolved stars (HOLMES; \citealt{Flores-Fajardo2011}), HII regions (therefore star formation), or a mixture of both effects. This scheme is simple, since it involves the usage of W$_{\rm H\alpha}$ only, but effective, since its predictions are largely consistent with BPT diagram ones. However, it does not aim to put a galaxy in a given category, but rather to classify galactic regions in diffuse ionised gas classes. Furthermore, it does not allow disentangling of excitation due to AGNs. 
Recent works of \cite{Sanchez2018} and \cite{Lacerda2020} define and explore the properties of AGN hosts from large samples of MaNGA and CALIFA galaxies, respectively, through classification based on the information from the BPT-diagrams and the W$_{\rm H\alpha}$ values of the central galaxy regions.

In this article, we synthesise the W$_{\rm H\alpha}$-values and BPT-diagram classification approaches in a similar way in order to propose a new bi-dimensional method that distinguishes between several "star formation quenching stages", considering the nuclear activity of the galaxy. The novelty of our method is that it defines galaxy groups not only based on the global W$_{\rm H\alpha}$ value or the W$_{\rm H\alpha}$ value at particular radii, but it considers the patterns described by the resolved W$_{\rm H\alpha}$ map across the entire galaxy. Our method is inspired by the previous work of \cite{Sanchez2013}, who first proposed to use the resolved W$_{\rm H\alpha}$ map to distinguish between SF and retired regions in the galaxies, extending the selection introduced by the WHAN diagram (see also \citealt{Sanchez2014}; \citealt{Sanchez-Menguiano2016,Sanchez-Menguiano2018}; \citealt{Sanchez2018}; \citealt{Cano-Diaz2016,Cano-Diaz2019}; \citealt{Lopez-Coba2019}; \citealt{Sanchez2020}). The proposed classification compares the properties of active and non-active systems with various ionisation distributions and aims to investigate the quenching mechanisms that transform galaxies from one type to another.

This study is organised as follows: Section \ref{S:data} describes the sample selection and the analysed data; Section \ref{S:2DELC} defines the criteria of the emission-line classification; Section \ref{S:EWprofiles} and \ref{S:ELCprop}  present the resolved and global properties of the EL galaxies.  Finally,  Section \ref{S:disc} and Section \ref{S:summary} present our discussion and summary, respectively.

\section{Data and analysis}
\label{S:data}
\subsection{Sample selection}
\label{SS:sample}
Our study is based on CALIFA integral-field unit (IFU) data, which observed 667 galaxies (in its 3$^{\rm rd}$ data release, \citealt{Sanchez2016c}) in the redshift range $0.005<z<0.03$ using the Potsdam Multi-Aperture Spectrophotometer (PMAS, \citealt{Roth2005}) in PPaK  (PMAS fiber Package, \citealt{Verheijen2004}, \citealt{Kelz2006}) mode, installed on the 3.5 m telescope at the Calar Alto observatory. In particular, CALIFA targets span various morphologies and stellar masses, being a representative sample of nearby Universe galaxies. Further details can be found in the survey-presentation paper (\citealt{Sanchez2012}), and the data-release papers (\citealt{Husemann2013}, \citealt{Garcia-Benito2015}, \citealt{Sanchez2016c}), and the sample paper of \cite{Walcher2014}.

For this work, we adopt the sample of 238 CALIFA galaxies originally explored in \citet[][hereafter \citetalias{Kalinova2017b}]{Kalinova2017b}. 
It was originally selected from \cite{Falcon-Barroso2017}, who provided stellar kinematics maps for 300 CALIFA galaxies, observed until June 2014. After exclusion of mergers and galaxies with unreliable dynamical models, \citetalias{Kalinova2017b} present a sample of 238 targets in a broad range of stellar masses (from 6$\times$$10^8$ M$_{\odot}$ to 5$\times$$10^{11}$ M$_{\odot}$) and types (from E1 to Sdm). The current sample is representative of the CALIFA mother sample (see Fig. 1 in \citetalias{Kalinova2017b}) and is therefore representative of the local Universe galaxy population too.

\subsection{Emission-line data}
\label{SS:ionisedgas}

There are three set of CALIFA data - low spectral resolution mode V500 (FWHM $\sim$6$ \AA$), high-resolution  V1200 (FWHM $\sim$2.3$\AA$), and a combination of both observation setups, called COMBO data (FWHM $\sim$6$\AA$). V500 ranges between 3745-7500~$\AA$, V1200 between 3650-4840~$\AA$, and COMBO between 3700-7500~$\AA$ \citep[see ][]{Sanchez2016c}.
Depending on scientific goals, we use different data-set in order to optimise our results. For example, COMBO dataset has the same resolution of V500, but it gives more reliable spectrophotometry on the blue end of its spectra  ($\lambda <$ 4600 {\AA}),  which contains important stellar population tracers (i.e., COMBO data are not influenced by vignetting effects).
For this set of galaxies, we make use of the ionised gas maps of H$\alpha$, H$\beta$, [NII], [SII], [OI], [OIII] line fluxes and equivalent width of H$\alpha$ (W$_{\rm H\alpha}$) obtained with the Pipe3D pipeline, using the low-resolution mode V500 of CALIFA data-set, where the lines are strong  and  present  in  this  spectral  window (in contrast to V1200 data), and available for all galaxies of our sample (in contrast to COMBO data). 

Full details about Pipe3D are given in \cite{Sanchez2016a,Sanchez2016b}, here we provide a short description. The pipeline uses combination of synthetic simple stellar populations (SSP) spectra from GRANADA library \citep{martins2005} and empirical ones from MILES project (\citealt{Sanchez-Blazquez2006}, \citealt{Vazdekis2010}, \citealt{Falcon-Barroso2011}). 
This joint SSP-library (see more details in \citealt{Cid-Fernandes2013}) adopts Salpeter  (\citealt{Salpeter1955}) initial mass function (IMF). It consists of 156 templates, covering 39 stellar ages (from 1 Myr to 14 Gyr), and 4 metallicities ranges (Z/Z$_{\odot}$ = 0.2, 0.4, 1 and 1.5). The library uses \cite{Girardi2000} evolutionary tracks, except for the youngest ages between 1 Myr and 3 Myr, which are based on Geneva tracks (\citealt{Schaller1992}; \citealt{Schaerer1993a,Schaerer1993b}; \citealt{Charbonnel1993}).

The original V-band data-cube is binned to increase the signal-to-noise, but keeping to minimum the overlap between regions with different physics. A spaxel-wise stellar population fit is performed using Fit3D. Afterwards, the model is evaluated by considering the continuum flux within each spatial bin. The ionised emission line data-cube is then constructed by subtracting the stellar-population model spaxels by spaxel.  Pipe3D extracts the main properties of 52 emission-lines, comprising the flux intensity, velocity, velocity dispersion and equivalent width maps generated using moment analysis techniques, alongside with their respective uncertainty maps.

\begin{figure*}[h!]
\centering
\includegraphics[width=0.65\textwidth]{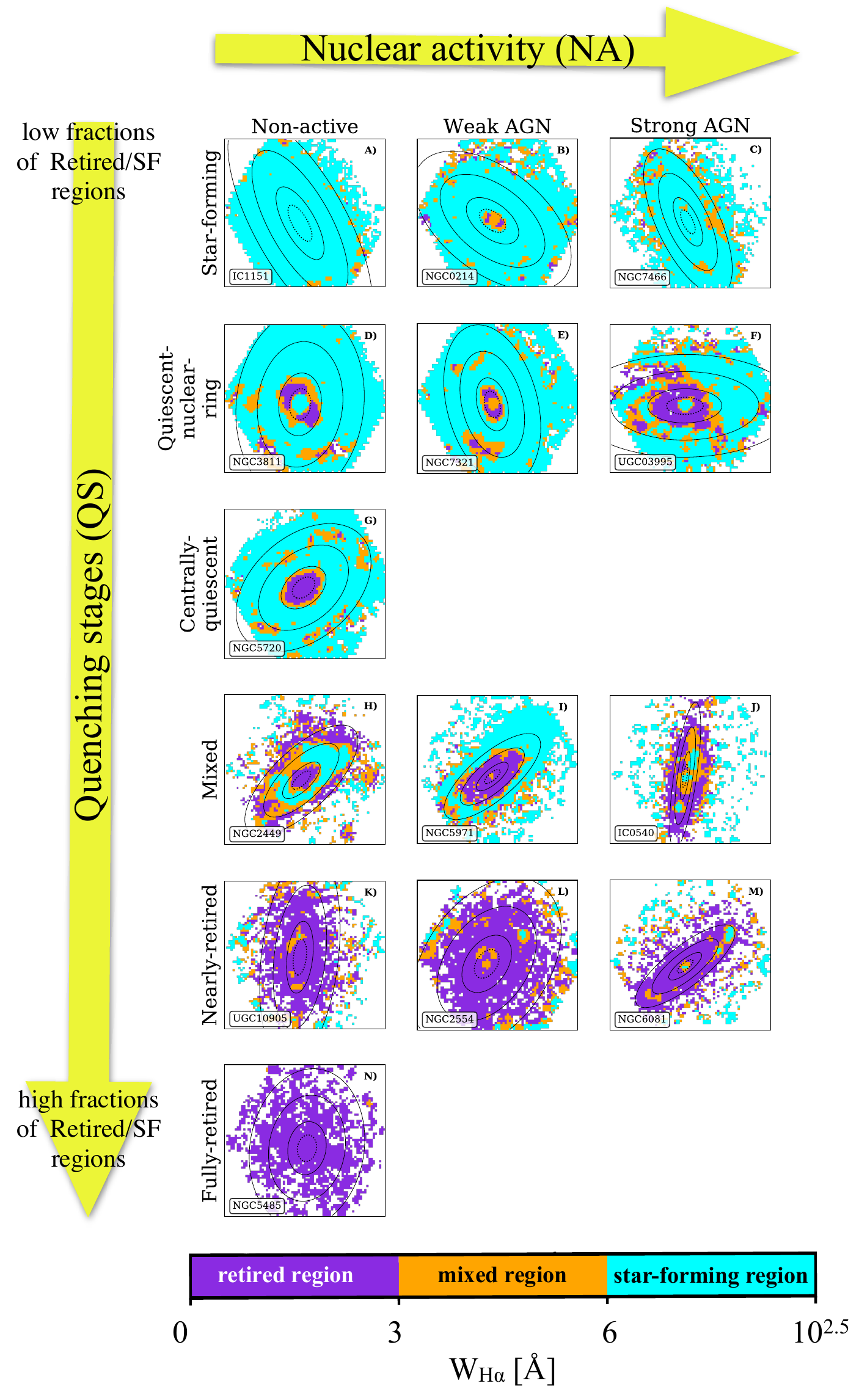}
\caption{2D Emission line classification "QuestNA" of our sample of galaxies (see Sec.\ref{S:2DELC}). The rows represent the quenching stages of the galaxies:  star-forming, quiescent-nuclear-ring, centrally-quiescent, mixed, nearby-retired and fully-retired with increasing "Retired regions/SF regions" ratio of the galaxy from the top to the bottom. The columns reflect the level of galaxy nuclear activity: non-active, weak and strong AGN from left to the right. 
Each panel of the scheme shows the W$_{\rm H\alpha}$ map of the galaxy, where the dashed contour indicates 0.5 of the effective radius, and the continuous contours correspond to 1, 2 and 3 $R_e$, respectively. The colour-bar of the W$_{\rm H\alpha}$ maps divides the regions of the galaxy on star-forming (W$_{\rm H\alpha}$ > 6 {\AA}), retired (W$_{\rm H\alpha} \leq$ 3 {\AA}) and mixed  (3 {\AA} < W$_{\rm H\alpha} \leq $  6 {\AA}).  
The emission-line classes of QuestNA are fourteen, instead of eighteen, because two of the quenching stages  centrally-quiescent and fully-retired are only non-active galaxies, which by definition  are fully retired in the central regions.}
\label{fig:2DELC}
\end{figure*}

\subsection{Stellar population data}
\label{SS:properties_pop}
For deriving star-formation history properties of the galaxies and optimising our results, we rely on COMBO dataset that provides a free vignetted FoV (field of view) on the blue side of the spectral range in contrast to V500  alone (although COMBO data is not available for 7 galaxies in our sample; see Table \ref{tab:long}). The stellar surface-mass-density ($\mu_*$), stellar mass ($M_*$), age ($\tau$), metallicity (Z) and star formation rate (SFR) of the galaxies has been estimated assuming a Chabrier IMF (\citealt{Chabrier2003}), which is necessary for the comparison of our sample  with Sloan Digital Sky Survey (SDSS\footnote{https://www.sdss.org/}) reference galaxies in the context of SFR-M* diagram exploration in Sec. \ref{S:SFRM}. 
Since {\sc PIPE3D} only adopts a Salpeter IMF, this makes it inconvenient for our analysis here. We run instead the {\sc STARLIGHT} spectral synthesis code of \cite{CidFernandes2005} on the COMBO data, fitting the individual spectra for each spatially resolved element of the IFU, and then implementing the output results in the platform of Python CALIFA {\sc STARLIGHT} Synthesis organiser (PyCASSO; \citealt{deAmorim2017}) to obtain the spatially resolved data.

For the fits, we use the CBe base of Charlot \& Bruzual (2007, priv. comm.), which is combination of 246 single stellar populations (SSP) spectra, covering 41 ages from 0.001 to 14 Gyr, and six metallicity ranges: $\log$ Z/Z$_{\odot}$=-2.3, -1.7, -0.7, -0.4, 0, +0.4. The SSP models are revised from \cite{Bruzual2003}, where both  MILES (\citealt{Sanchez-Blazquez2006}, \citealt{Falcon-Barroso2011}) and GRANADA (\citealt{martins2005}) stellar libraries replace the original library STELIB (\citealt{LeBorgne2003}). Further,  "Padova 1994" stellar tracks has been applied (\citealt{Alongi1993}; \citealt{Bressan1993}; \citealt{Fagotto1994-III,Fagotto1994-IV,Fagotto1994-V}; \citealt{Girardi1996}). More details about the SSP analysis, we refer to related works of  \cite{Cid-Fernandes2013}, \cite{Gonzalez-Delgado2015}, \cite{deAmorim2017}, \cite{Garcia-Benito2017,Garcia-Benito2019}. \\

The mean stellar surface brightness, $\mu_*$, is extinction corrected and calculated within 1 $R_e$. The galaxy age ($\tau$) and metallicity ($Z$) are light- and mass-weighted within 1 $R_e$, respectively.  The total star formation rate is calculated from the spatially resolved star formation history of galaxies by adding the amount of stellar mass formed into stars in the last 32 Myr (and divided by this time interval; \citealt{Gonzalez-Delgado2016,Gonzalez-Delgado2017,deAmorim2017}). This approach helps to account for the star-formation of the galaxies that is heavily obscured or with low quantity of ionised gas (i.e. those galaxies where the H$_\alpha$ emission is very weak). 
The SFR from H$_\alpha$ has a shorter timescale of emission in comparison to the SFR from SSP, which mainly comes from OB stars (3 Myr, \citealt{Kennicutt_Evans2012}). In addition, the SSP based approach averages over stochastic star formation to provide a good estimation of the SFR.

\subsection{Photometric data}
\label{SS:properties_phot}
The photometric properties of our galaxies are analysed using SDSS data (\citealt{Abazajian2009}, \citealt{Alam2015}).
The bulge-to-disk (B/D) ratios of the galaxies is derived from their bulge-to-total (B/T) and disk-to-total (D/T) flux ratios, provided by \cite{Mendez-Abreu2017}, who analyse $r$-band  SDSS images from Data Release (DR) 7 (\citealt{Abazajian2009}) of 404 CALIFA galaxies. The overlap with our sample (for galaxies with both available B/T and D/T ratios) is 127 galaxies, which approximately covers half of the galaxies within each EL class, with no significant changes across our classes. 
Thus, we do not expect large bias in our conclusions regarding B/D ratios.

The total luminosity (L$^{\mathrm{tot}}_{\mathrm{r}}$) of the galaxies is calculated from the $r$-band SDSS (DR12; \citealt{Alam2015}) images via Multi-Gaussian Expansion method (MGE; \citealt{Emsellem1994}) by  \citetalias{Kalinova2017b}.

Effective radii ($R_e$) are measured via growth curve analysis using elliptical apertures on SDSS DR7 (\citealt{Abazajian2009}) images, which provides reliable estimations in case of highly inclined galaxies (Sanchez at al., in prep). The galaxy morphology and the presence of bar are defined visually by few members of the CALIFA team as described in \cite{Walcher2014}.

\subsection{Kinematical and dynamical data}
\label{SS:Kindyn_prop} 
To calculate the galactic specific angular momentum ($\lambda_{R_e}$) of our galaxies, we use the public available stellar kinematic maps of the CALIFA galaxies, provided by \cite{Falcon-Barroso2017}. They use the high-resolution V1200 setup of CALIFA data, 
which has much higher spectral resolution ($\sim$ 72 \kms) than V500 grating ($\sim$ 139 \kms), and where V500 has difficulties to reliably measure velocity dispersions below 100 \kms (see Sec. 4.2 of their paper).  
Further, \cite{Falcon-Barroso2017} derive the line-of-sight velocity ($V$) and velocity dispersion ($\sigma$) distributions after fitting the stellar continuum of the galaxies in each spatial bin using the \texttt{pPXF} code of \cite{Cappellari2004} with stellar templates from the Indo-US spectral library (\citealt{Valdes2004}). The averaging of the pixels has been done through the Voronoi 2D binning method of \cite{Cappellari2003} with a minimal signal-to-noise ratio of 20 per pixel.

We calculate the specific angular momentum $\lambda_{R_e}$ using equation (6) in \cite{Emsellem2007}: 
\begin{equation}
\lambda_{\mathrm{R_e}} = \frac{\sum_{i=0}^{N_k}F_iR_i|V_i|}{\sum_{i=0}^{N_k}F_iR_i\sqrt{V^2_i+\sigma^2_i}} 
\label{eq:lambdaR}
\end{equation}
where $F_i$,$R_i$,$V_i$ and $\sigma_i$ are the flux, circular radius, velocity and
velocity dispersion of the $i$-th spatial bin. 

We report $\lambda_{R_e}$ as the integrated value within one effective radius of the galaxy, reached by all of our targets. Our measurements of $\lambda_{R_e}$ are comparable to the published calculations of \cite{Falcon-Barroso2019}, although we adopt different values for the position angles and the inclinations of the galaxies (from \citetalias{Kalinova2017b}).

The amplitude (V$_{\mathrm{c,max}}$) of the circular velocity curve of the galaxies within 1.5$R_e$ is inferred from stellar dynamical modelling (Sec. 3.1 in \citetalias{Kalinova2017b}) based on axisymmetric Jeans equations 
(i.e., using JAM\footnote{http://purl.org/cappellari/software} code of \citealt{Cappellari2008}).

In addition, the total dynamical mass M$^{\mathrm{tot}}_{\mathrm{dyn}}$ (panel K) comes from the multiplication between the total luminosity (L$^{\mathrm{tot}}_r$), derived via MGE method (\citealt{Emsellem1994}), and the dynamical mass-to-light 
ratio $\gamma_{\mathrm{dyn}}$ of the galaxies, obtained via Markov Chain Monte Carlo approach (MCMC; 
\citealt{Foreman-Mackey2012}) of JAM model  (see \citealt{Kalinova2017a}). Our masses are comparable to the recent calculations of the dynamical mass by \cite{Aquino-Ortiz2018}, who use the kinematic parameter S$^2_k =K\,V_{rot}^2+\sigma^2$ as a dynamical proxy (where $K$ is constant, V$_{rot}$ is the rotation velocity, and $\sigma$ is the velocity dispersion of the galaxy; \citealt{Weiner2006}). To calibrate the constant $K$, they use the dynamical mass measurements of \cite{Leung2018}, \cite{Zhu2018a}, and \cite{Zhu2018b}.  

Through M$^{\mathrm{tot}}_{\mathrm{dyn}}$ and M$^{\mathrm{tot}}_{\mathrm{*}}$, we calculate the mass discrepancy factor $f_d=1-(\mathrm{M^{tot}_*/M^{tot}_{dyn}})$, showing the relationship between the total stellar and dynamical masses as a fraction.

\section{2D Emission-line classification}
\label{S:2DELC}

\begin{figure*}
\centering
\includegraphics[width=0.9\textwidth]{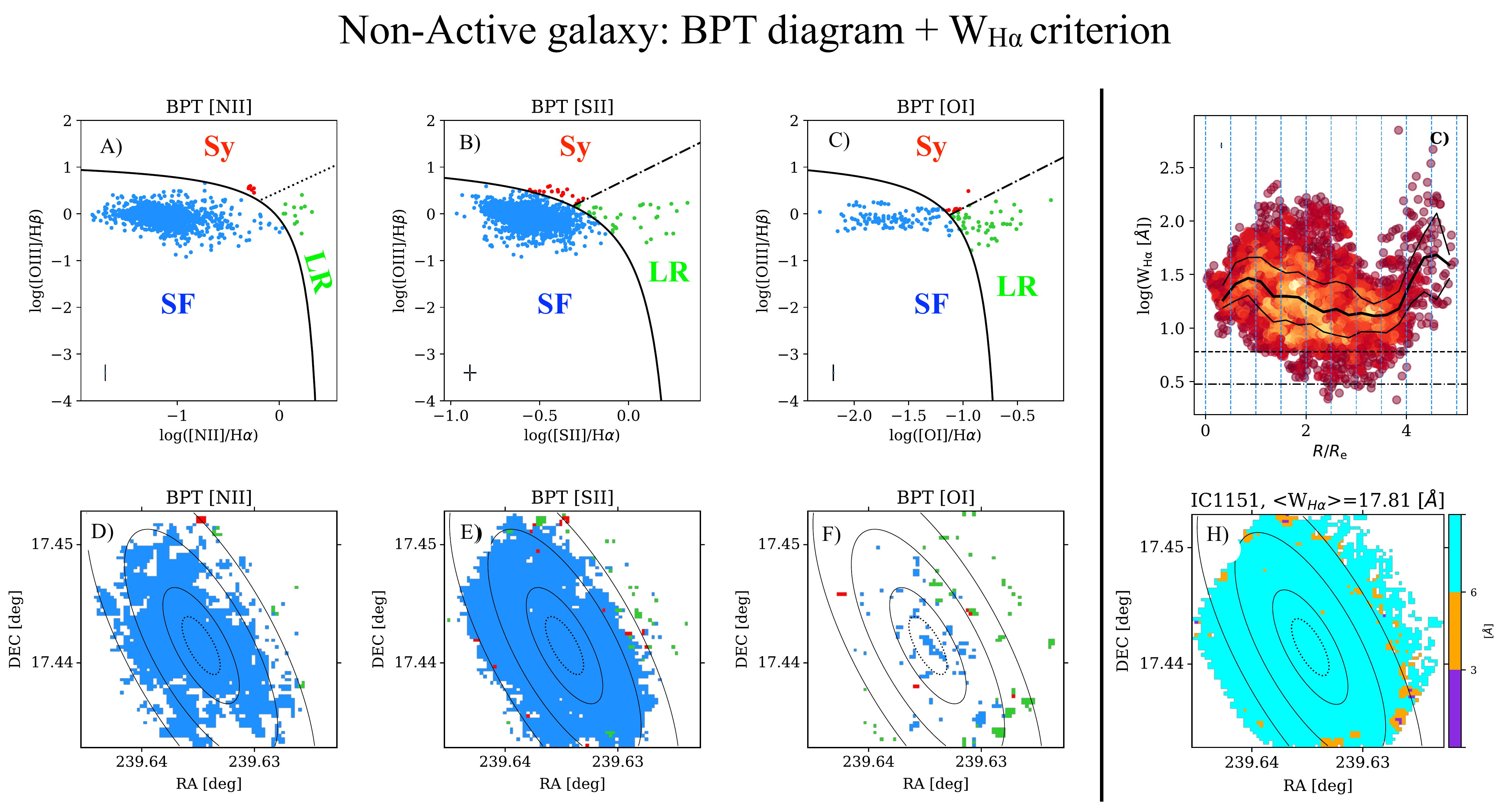}
\caption{Galaxy IC1151 as example of non-active class based on BPT diagram and W$_{\rm H\alpha}$ criterion  (see Sec. \ref{S:2DELC}).
\emph{Panels A, B, and C:} Resolved [NII]$-$,[SII]$-$ and [OI]$-$BPT diagnostic diagrams, where the solid curve is the empirical SF line of \citet{Kauffmann2003} and pixels below this line are dominated by star-formation (SF; blue). The division (dash-dotted line) between Seyfert (Sy; red) and LINER (LR; green) region in panels B and C is adopted from \citet{Kewley2006}, while the empirical Seyferts/LINER devision (dotted line) in panel A comes from \citet{CidFernandes2010}. \emph{Panels D, E, and F:} The corresponding to the BPT diagram resolved [NII]$-$,[SII]$-$ and [OI]$-$BPT maps. \emph{Panels G:} Resolved  W$_{\rm H\alpha}$ radial profile from individual spaxels, where the dashed-dotted and the dashed lines correspond to the thresholds of 3 and 6 {\AA}, respectively. The blue dotted vertical lines define the bins of 0.5 R$_e$. The black thick line shows the median value of the W$_{\rm H\alpha}$, where the thin black lines indicate the median absolute deviation. The morphological type, QS and the nuclear activity of the galaxy are provided as labels in the panel.
\emph{Panel H:}  Resolved  W$_{\rm H\alpha}$ map, where the colour bar indicates the thresholds of the spaxels: W$_{\rm H\alpha}$ $\leq$ 3 {\AA} (violet), 
3 {\AA}  < W$_{\rm H\alpha}$ $\leq$ 6 {\AA} (orange) and W$_{\rm H\alpha}$ > 6 {\AA} (cyan).
The dotted line indicates the region of 0.5 R$_e$, where the continuous lines mark separation of 1, 2 and 3 R$_e$.  The median value of the W$_{\rm H\alpha}$ for the full maps is given as label in the panel.
To classify this galaxy as non-active, the central pixels that are below 0.5 $R_e$ should not populate the Seyfert (Sy) region of the BPT diagrams (at least  two from the three BPT diagrams), and their W$_{\rm H\alpha}$ values could vary, depending on the quenching stage of the galaxy. }
\label{fig:NonA}
\end{figure*}

\begin{figure*}
\centering
\includegraphics[width=0.9\textwidth]{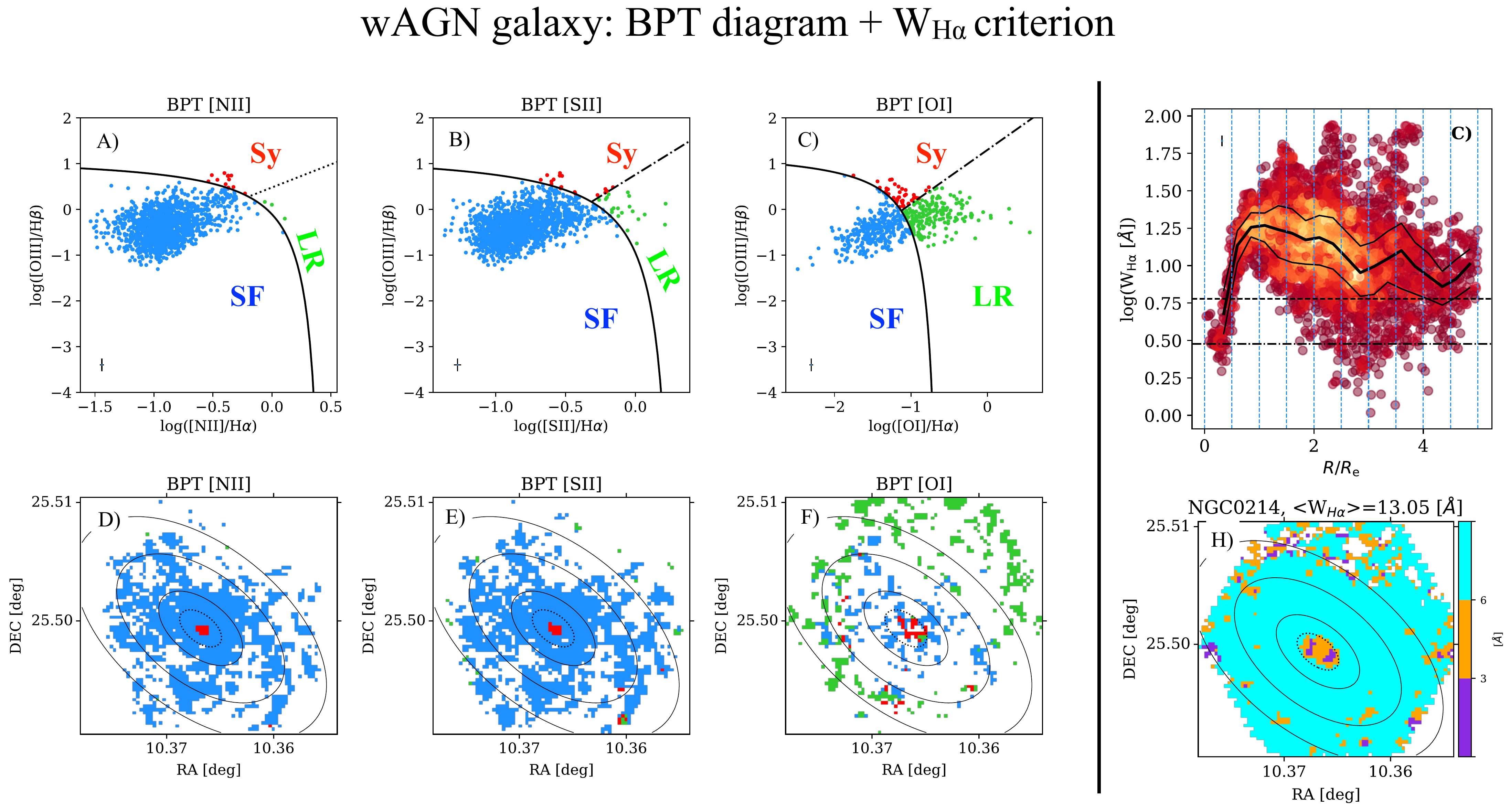}
\caption{Galaxy NGC0214 as example of weak AGN class based on BPT diagram and W$_{\rm H\alpha}$ criterion (see Sec. \ref{S:2DELC}). Description as per Fig. \ref{fig:NonA}.
To classify this galaxy as wAGN, the central pixels that are below 0.5 $R_e$ should have values 3 {\AA} < W$_{\rm H\alpha}$ $\leq$ 6 {\AA}, as well they should populate the Seyfert (Sy) region of the BPT diagrams (at least two from the three BPTs). }
\label{fig:wAGN}
\end{figure*}

\begin{figure*}
\centering
\includegraphics[width=0.9\textwidth]{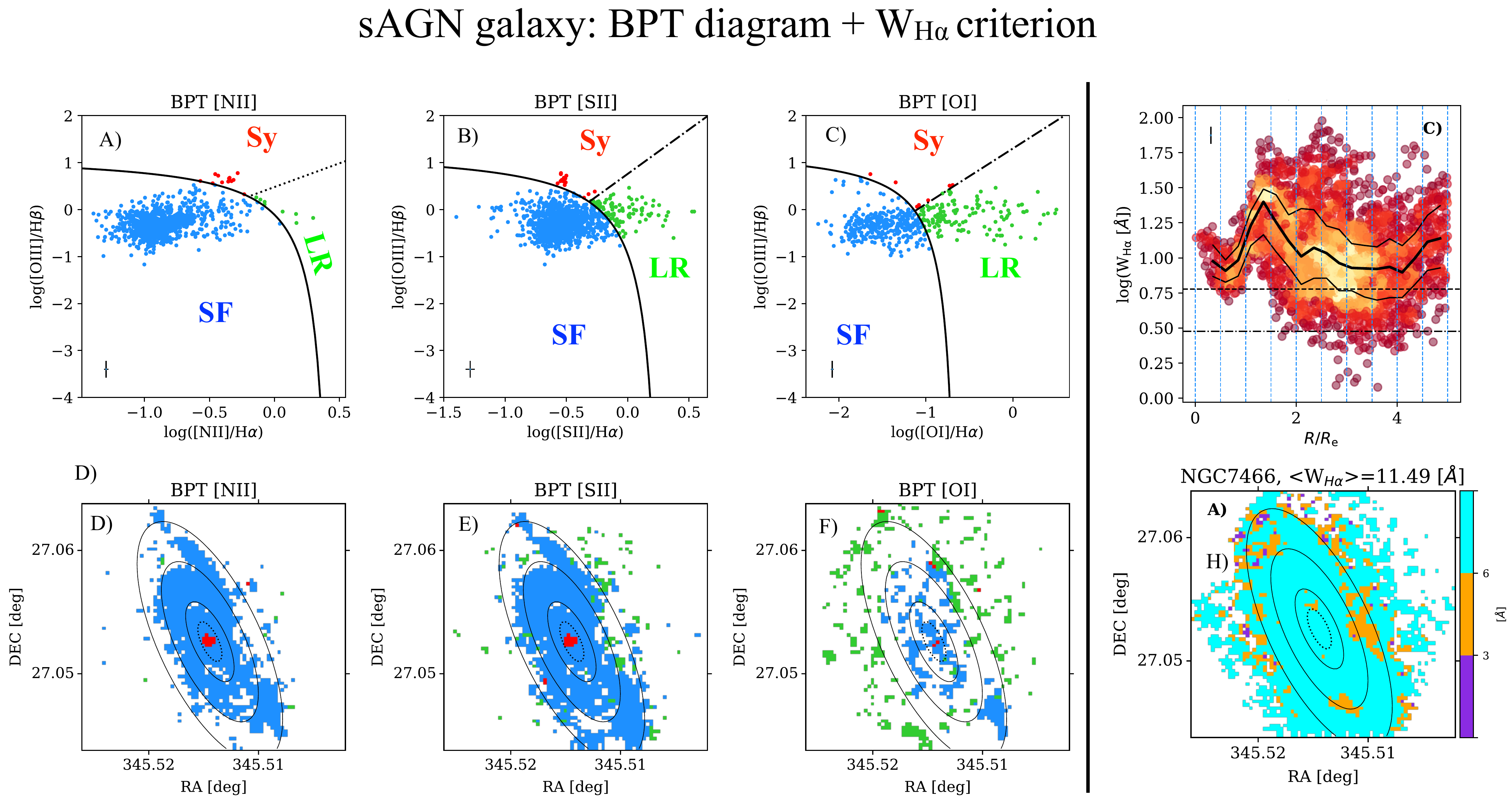}
\caption{Galaxy NGC7566 as example of strong AGN class based on BPT diagram and W$_{\rm H\alpha}$ criterion (see Sec. \ref{S:2DELC}). Description as per Fig. \ref{fig:NonA}.
To classify this galaxy as sAGN, the central pixels that are below 0.5 $R_e$ should have W$_{\rm H\alpha}$ values above 6 {\AA} and they should also populate the Seyfert (Sy) region of the BPT diagrams (at least two from the three BPTs).}
\label{fig:sAGN}
\end{figure*}

We classify galaxies according to their ionisation, following recent studies in the field 
(\citealt{Schawinski2007}, \citealt{Stasinska2008},\citealt{CidFernandes2011},\citealt{Singh2013}, \citealt{Belfiore2016}, \citealt{Sanchez2018}, \citealt{Sanchez2020}, \citealt{Lacerda2020}). In particular, our classification is based on the peculiar patterns defined by the spatially resolved H$\alpha$-equivalent-width (W$_{\rm H\alpha}$) value within the galaxies (similar to \citealt{CidFernandes2011}, \citealt{Sanchez2018}, 
\citealt{Lacerda2018}, \citealt{Lacerda2020}, \citealt{Espinosa-Ponce2020}). In essence, star-forming or HII regions show $W_{\rm H\alpha}>$ 6 {\AA}; finding $W_{\rm H\alpha} \leq$ 3 {\AA} indicates that the gas in the region is ionised by the old stellar population (post-AGB stars or HOLMES); values of 3 {\AA} $<W_{\rm H\alpha} \leq$ 6 {\AA} are consistent neither with star formation nor HOLMES, but rather with other mixed phenomena (e.g., diffuse ionised gas, high-velocity shocks; e.g., \citealt{Lacerda2018}). Therefore the $W_{\rm H\alpha}$ values of 3 {\AA} and 6 {\AA} mark the thresholds we use to recognise patterns in the resolved $W_{\rm H\alpha}$ map (see also \citealt{CidFernandes2011}).

To distinguish between the active and non-active galaxies of our sample (i.e., whether or not they host an AGN in their centres), we use three BPT excitation diagnostic diagrams (and their corresponding maps) involving the [OIII], [SII], [OI] and [NII] line ratios with respect to the Balmer lines (\citealt{Baldwin1981}, \citealt{Veilleux1987}). We assess these diagrams using the loci separating different ionisation regimes proposed by \citet{Kauffmann2003}. The utility of combining W$_{\rm H\alpha}$ together with the BPT diagnostic diagrams is suggested by many authors (e.g., \citealt{Sarzi2010}, \citealt{CidFernandes2010}, \citealt{Singh2013}, \citealt{Sanchez2014}, \citealt{Lacerda2018}, \citealt{Lacerda2020}). Thus, the galaxy is defined as a host of a weak AGN if its central pixels, located within 0.5 R$_e$ of the galaxy, populate the Seyfert region of the BPT diagrams and their W$_{\rm H\alpha}$ values ranges between 3 {\AA} and 6 {\AA}. In case the W$_{\rm H\alpha}$ values of those central pixels go above 6 {\AA}, then the galaxy is considered a host of a strong AGN (see Sec. \ref{SS:NA_criteria} for more details.)

We apply a 1-$\sigma$ clipping to the W$_{\rm H\alpha}$ maps and measurements in the BPT-diagrams, which removes low signal-to-noise spaxels.  We experimented with a stricter 3-$\sigma$ clipping, but this omitted targets with lower signal-to-noise in the emission-line data (typically for old population galaxies).

Visually identifying dominant patterns in the $W_{\rm H\alpha}$ maps of CALIFA galaxies, we find six quenching stages (QS) -  star-forming, quiescent-nuclear-ring, centrally-quiescent, mixed, nearly-retired and fully-retired. On the other hand, from the BPT diagrams we distinguish three nuclear activities (NA) of the galaxies - non-active (NonA), weak AGN (wAGN) and strong AGN (sAGN).

If we explore the properties of the galaxies only by the QS criteria, we refer to a "quenching classification", while if we explore the galaxies only through their nuclear activity criteria, we link to a "nuclear activity classification". 
In Fig. \ref{fig:2DELC}, we combine both QS and NA criteria and define a two-dimensional (2D) emission-line classification called \emph{QuestNA} (short from QUEnching STages and Nuclear Activity).

\emph{QuestNA} can be expressed as a matrix of six QS rows by three NA columns that gives eighteen EL classes, but in reality the EL classes only show fourteen populated classes, because centrally-quiescent and fully-retired QSs do not allow a nuclear activity in the centre of the galaxies by definition $-$ the $W_{\rm H\alpha}$ values in these regions is equal to or below 3 {\AA}, which is too low to distinguish the AGN activity from the other processes in the galaxies (e.g.,  \citealt{CidFernandes2010},  \citealt{Sanchez2014}, \citealt{Lacerda2020}). The QS rows are ordered by increasing of the retired regions in the galaxy in comparison to the star-forming ones, while the NA columns are organised by the increasing of the nuclear activity of the galaxy (i.e., nonA$\rightarrow$wAGN$\rightarrow$sAGN).

\subsection{Quenching stage (QS) criteria}
\label{SS:QS_criteria}
Based on the distribution of ionisation in different regions within the 238 CALIFA galaxies and visual inspection of their W$_{\rm H\alpha}$ maps, we propose the following quenching stages of the galaxies (i.e., this is the first dimension of the emission-line classification):

(i) \emph{Star-forming galaxies:} 
First row of panels (A$-$C) of  Fig. \ref{fig:2DELC} exemplifies the typical star-forming class galaxies (NGC1151, NGC0214 and NGC7466) from each type of nuclear activity (nonA, wAGN and sAGN, respectively). The median value of the W$_{\rm H\alpha}$ spaxels should be larger than 6 {\AA} and there are no large regions in these galaxies with low-ionisation emission lines, where W$_{\rm H\alpha}<3$~{\AA} (such low ionisation spaxels are usually less than 3\% of the total ones). The star-forming class can be represented by both active and non-active galaxies.
 
 (ii) \emph{Quiescent-nuclear-ring galaxies:}
The second row of panels (D$-$F) of Fig \ref{fig:2DELC} represents the quiescent-nuclear-ring class galaxies (NGC3811, NGC7321 and UGC03995) from each type of nuclear activity. This group is characterised by a quiescent-nuclear-ring region within the 0.2$-$0.5 $R_e$ of the galaxy, where the W$_{\rm H\alpha}$ is below 3 {\AA}. On the other hand, the central spaxels, which are surrounded by the ring, always have  W$_{\rm H\alpha}$ values above 3 {\AA}. In the outer regions of the galaxies (0.5-2.0 $R_e$), there is an evidence for the presence of star-forming disk with W$_{\rm H\alpha}$ values, going above 6 {\AA}. The quiescent-nuclear-ring class contains both active and non-active galaxies.
 
(iii) \emph{Centrally-quiescent galaxies:}
In panel G of Fig. \ref{fig:2DELC}, galaxy NGC5720 represents the centrally-quiescent class galaxies, which shows low-ionisation EL regions at small galactocentric radii $-$ W$_{\rm H\alpha}$ stays below 3 {\AA} up to 0.5 $R_e$. There is an evidence for the presence of star-forming disk since  W$_{\rm H\alpha}$ values goes above 6 {\AA}  within the range 0.5$-$2.0 $R_e$. The centrally-quiescent class are only non-active galaxies since the W$_{\rm H\alpha}$ value of the central pixels is always equal or below 3 {\AA} even in case that the line ratios populate the Seyfert region of the BPT diagrams. This class is similar in essence to the central low-ionisation emission-line region (cLIER) class galaxies from the [SII]-BPT classification of \cite{Belfiore2016}.

(iv) \emph{Mixed galaxies:} 
The fourth row of panels (H, I and J) of Fig. \ref{fig:2DELC} exemplifies the mixed galaxies (NGC2449, NGC5971 and IC0540). This class is characterised by star-forming regions (high-ionisation EL regions, where W$_{\rm H\alpha}$ > 6 {\AA}), intermediate-ionisation EL regions (3 {\AA} <W$_{\rm H\alpha}$ $\leq$ 6 {\AA}) and low-ionisation emission-line regions (W$_{\rm H\alpha}$ $\leq$ 3 {\AA}) with no clear morphological pattern of the W$_{\rm H\alpha}$ map. The median value of W$_{\rm H\alpha}$ of the galaxy is usually between 3 and 6 {\AA}.  A low-ionisation EL region in the centre of the galaxies is possible, but it should be more extended than 0.5$R_e$ (to be distinguishable from the centrally-quiescent class) or randomly distributed within the disk. 
This mixed class can be represented by both active and non-active galaxies.

(v) \emph{Nearly-retired galaxies: }
In the fifth row of panels (K, L and M) of Fig. \ref{fig:2DELC} exemplifies the nearly-retired class galaxies (UGC10905, NGC2554 and NGC6081), where the median value of the W$_{\rm H\alpha}$ map is equal or below 3 {\AA}. Similar to the mixed class, nearly-retired galaxies has no clear pattern, but these galaxies are mostly dominated by the low-ionisation EL regions (i.e., above 90 per cent  of the spaxels usually have values of the W$_{\rm H\alpha}$ that is equal or below 3 {\AA}) and little star-formation  within the 2 $R_e$ of the galaxy (usually less than 3 per cent of the spaxels have W$_{\rm H\alpha}$ > 6 {\AA} ). This class can not be mistaken with the cQ class in case of presence of low-ionisation ELs in its central regions since it does not have a well defined star-forming disk beyond 0.5 $R_e$ as the centrally-quiescent class has. The nearly-retired class can be represented by both active and non-active galaxies.

(vi) \emph{Fully-retired galaxies:}
Panel N of Fig. \ref{fig:2DELC} represents the fully-retired class galaxies (NGC5485), where the median value of W$_{\rm H\alpha}\le 3$~{\AA}. The galaxy is fully dominated by these low-ionisation EL regions within the 2 $R_e$ of the galaxy. Similar to the centrally-quiescent class, fully-retired class are only non-active galaxies since the W$_{\rm H\alpha}$ value of the central pixels is always equal or below 3 {\AA} even in case that the line ratios populate the Seyfert region of the BPT diagrams. This class is somehow similar to the low-ionisation emission-line region (eLIER) galaxies in the [SII]-BPT classification of \cite{Belfiore2016}. 

\subsection{Nuclear activity (NA) criteria}
\label{SS:NA_criteria}
From the BPT diagrams in panels A$-$C of Figures \ref{fig:NonA} $-$ \ref{fig:sAGN}, we define if the galaxy is active or non-active for each quenching stage (i.e., this is the second dimension of the EL classification \emph{QuestNA}). If the line-ratios of the central spaxels that are below 0.5 $R_e$ populate the Seyfert region of the BPT diagrams (at least two from the three BPT diagrams), then the galaxy is defined as an AGN-host candidate (three is the minimum number of the central pixels that define the nuclear activity of the galaxies based on the Point Spread Function of CALIFA maps).

If the  central pixels (< 0.5 $R_e$) have values of the W$_{\rm H\alpha}$ above 3 {\AA}, the nuclear activity classification is reliable, and the galaxy is considered active, otherwise it is labelled non-active (galaxy is also classified as non-active if the line-ratios do not populate the Seyfert regions). Once the galaxy is classified as an active, we further define if it hosts a weak AGN, where the W$_{\rm H\alpha}$ values of the central pixels ranges between 3 and 6 {\AA}, or a strong AGN, where the W$_{\rm H\alpha}$ values of the central pixels go above 6  {\AA}.  The corresponding resolved maps of the BPT diagrams in panels D$-$F helps us to locate if the Seyfert spaxels belong to the centre of the galaxy. Panel G and H are showing the resolved radial profile and map of the W$_{\rm H\alpha}$, respectively, that are needed for applying the thresholds of the nuclear activity classification (i.e., 3 and 6 {\AA} of the W$_{\rm H\alpha}$). 

All resolved maps and BPT diagrams for the rest of the sample are shown in the online Appendix \ref{A:EL_examples}.

\subsection{Uncertainties of the EL classification}
\label{SS:uncertainties}
Using the scheme, we perform a by-eye classification of our galaxy sample, finding
96 star-forming, 12 quiescent-nuclear-ring, 23 centrally-quiescent, 37 mixed, 34 nearly-retired and 36 fully-retired galaxies (from which 215 nonA, 8 wAGN and 15 sAGN).
Despite of the detailed analysis, there are marginal cases in our classification,  where a galaxy could be slightly below or above certain thresholds of the median W$_{\rm H\alpha}$ and/or its radial profile. Thus, we attribute a quality index to our classified galaxies: a flag value of 1 indicates an unambiguous QS classification; flag=2 indicated uncertain QS classification due to complex W$_{\rm H\alpha}$ morphology; and flag=3 indicates uncertain QS classification due to bad data. Approximately 10\% of the galaxies have uncertain QS classification (flag=2 or 3). On the other hand, the quality index, related to the nuclear activity classification is called sure (S) or unsure (U) case, depending if the nuclear activity is confirmed by the three or two of the BPT diagrams, respectively (9 galaxies have label=U, which is $\sim$ 4\% from the total 238 galaxies and $\sim$ 40\% from the total active galaxies). The QS and NA classification labels, with the corresponding flags for all 238 CALIFA galaxies, are given in Table \ref{A:table}. 

Related to the AGN (weak and strong) classes, it is possible that the ionisation in the centre of the galaxy, shown in the Seyfert region of the BPT diagrams, is caused by SF-driven shocks rather than the AGN itself (e.g., \citealt{Lopez-Coba2019}). In this case, the galaxies typically show a bi-conical outflow from the centre of the galaxy to the outskirts, confirmed by the velocity-to-velocity dispersion ratio of the galaxy in the central regions. \cite{Lopez-Coba2019} examined 667 CALIFA galaxies from the Data Release 3 (\citealt{Sanchez2016c}) in which our sample is included and defined 17 galaxies with SF-driven outflows. We compare our sample with theirs, and we only exclude the galaxy NGC0681 from our list of the wAGN candidates due to the presence of outflow.

We also compare our AGN results with the list of \emph{bona fide} AGN host galaxies, defined by \cite{Lacerda2020} from the extended sample of CALIFA survey (\citealt{Sanchez2016c}, \citealt{Galbany2018}). There is an overlap of 10 galaxies
(eight galaxies in the sAGN class: UGC03995, MCG-02-02-030, UGC00987, NGC2410, IC2247, NGC2639, IC0540, NGC3160 and two galaxies in the wAGN class: NGC2554 and NGC6762) from the total 23 AGN (15 sAGN and 8 wAGN) host galaxies in our sample. We classify more galaxies as having AGN because of the more relaxed criteria we adopt in our study compared to \cite{Lacerda2020}.  Specifically, we require the Seyfert region of the BPT diagrams to be populated at least for two of the three diagrams but \citet{Lacerda2020} require population in all three diagrams. The goals of these two studies are different: \cite{Lacerda2020} focus on the \emph{bona fide} cases of AGN hosts and exclude some boarder cases, while we aim to record any possibility of AGN activity in the galaxy. For example, NGC1167, which has a strong AGN in the centre (e.g., \citealt{Struve2010}), is included in our AGN sample but would be excluded by the stricter requirements of \cite{Lacerda2020}.
Outflows and shocks are not included in the current 2D EL classification since it is out of the scope of the current study. On the other hand, it might be possible to add their categorisation to the nuclear activity criteria and expand the EL classes.

Finally, we might expect to find more EL classes or sub-classes with improved spectral and spatial resolution data, and a higher number of sample statistics in comparison to the current sample (e.g., the AMUSING ++ survey, \citealt{Lopez-Coba2020}).

\section{Characteristic W$_{\rm H\alpha}$ radial profiles of the EL galaxies}
\label{S:EWprofiles}
\begin{figure*}
\centering
\includegraphics[width=0.95\textwidth]{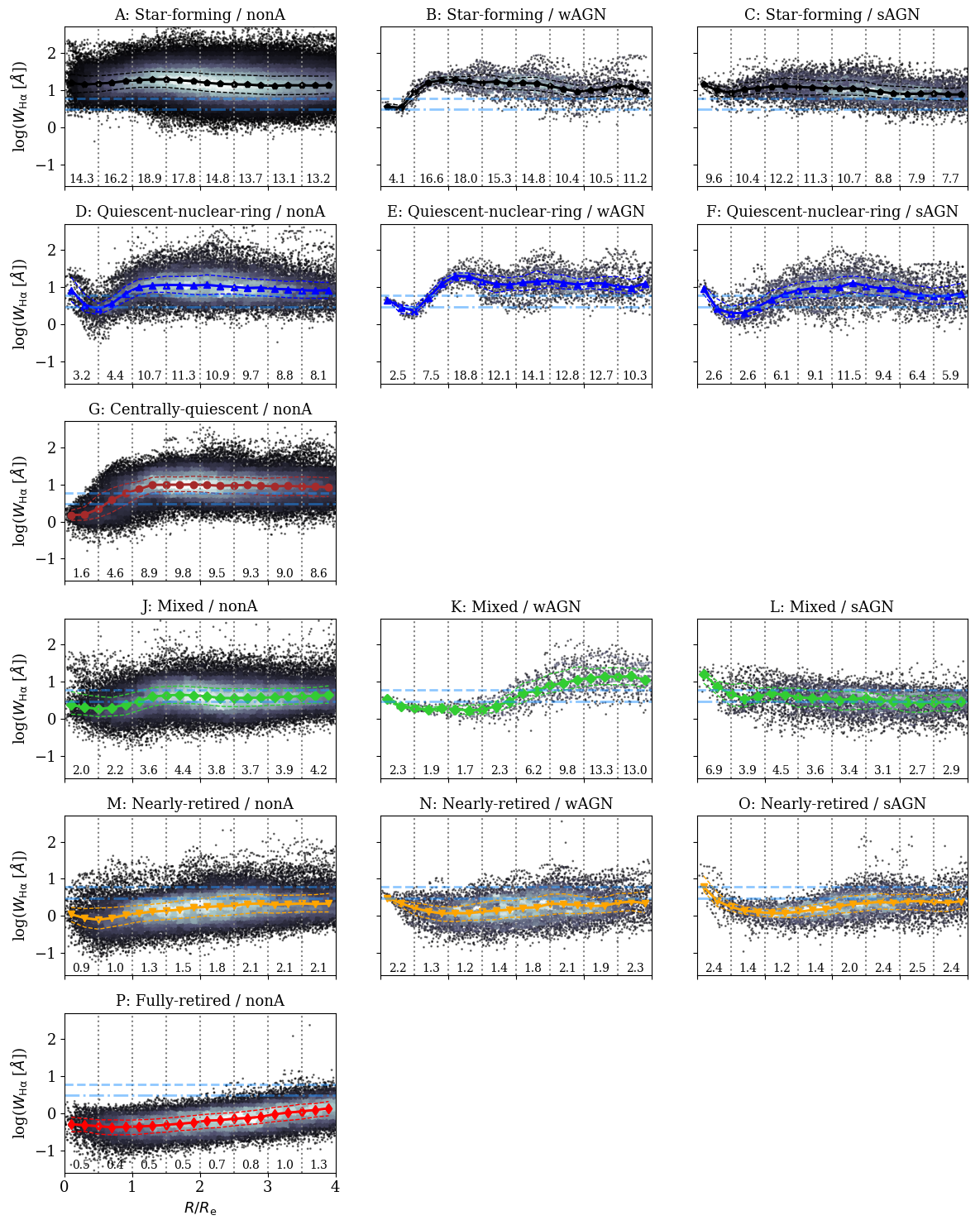}
\caption{Stacked radial profiles equivalent width of H$\alpha$ (W$_{\rm H\alpha}$) for the galaxies in the different emission-line classes, normalised by the galaxies effective radius (Sec. \ref{S:EWprofiles}). In the panel B to I, the bi-dimensional histograms indicate the density of point at particular $W_{\rm H\alpha}$ and $R/R_{\rm e}$ values. The coloured full lines show the median of the radial distributions, the dashed colour lines indicate the $25^{\rm th}$ and $75^{\rm th}$ percentiles of the radial distributions. The blue dash-dotted and dashed horizontal lines display constant $W_{\rm H\alpha}=3$ {\AA} and $W_{\rm H\alpha}=6$ {\AA} values, respectively. The numbers on the bottom of each panel show the median $W_{\rm H\alpha}$ in a given radial bin of $0.5\,R/R_{\rm e}$. Bins are separated by vertical, black, dashed lines. }
\label{fig:EWprof_matrix}
\end{figure*}

\begin{figure*}
\centering
\includegraphics[width=1.0\textwidth]{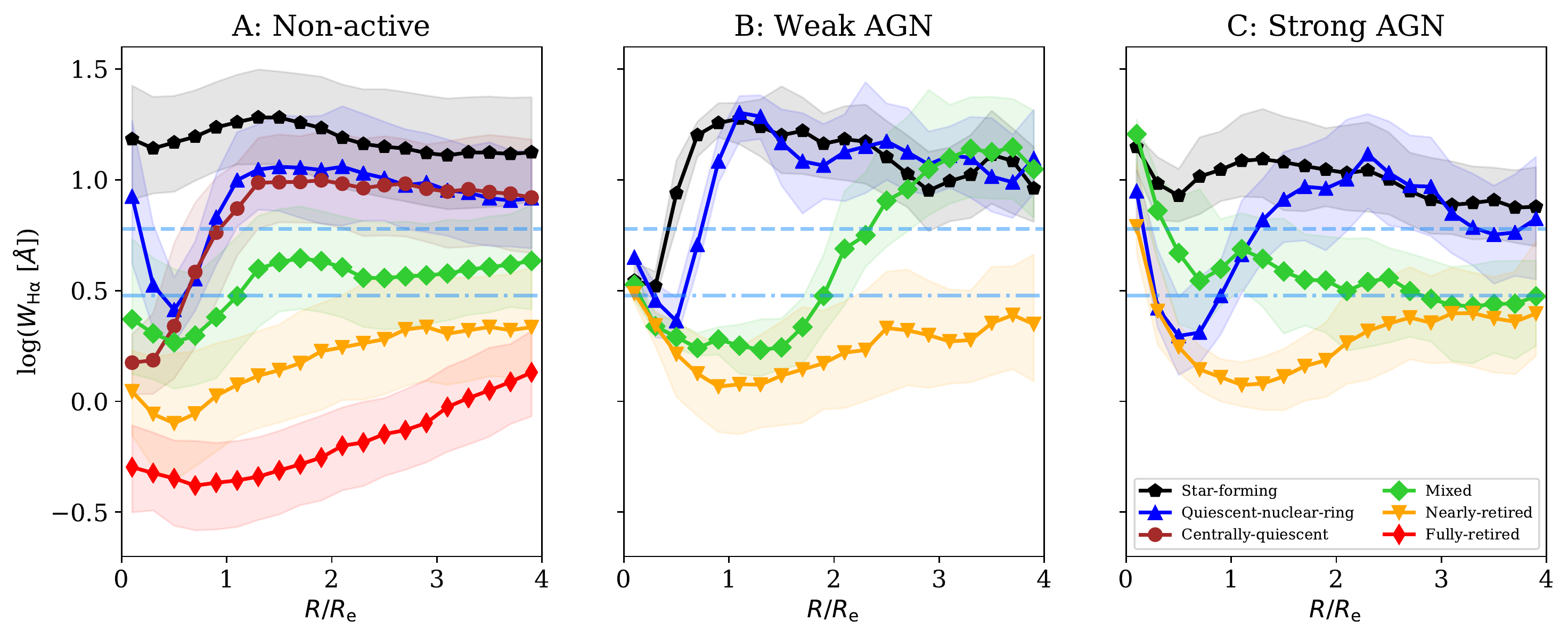}
\caption{The median radial profiles equivalent width of H$\alpha$ (W$_{\rm H\alpha}$) for our sample of galaxies in each quenching stage for non-active (left), wAGN (middle) and sAGN (right) classes, normalised by the galaxies effective radius (Sec. \ref{S:EWprofiles}). The shaded regions express the $25^{\rm th}$ and $75^{\rm th}$ percentiles of the radial distributions. The blue dash-dotted and dashed  horizontal lines display $W_{\rm H\alpha}=3$ {\AA} and $W_{\rm H\alpha}=6$ {\AA} thresholds, respectively. The W$_{\rm H\alpha}$ profiles' variation suggests an "inside-out" star-formation quenching scenario for our sample of galaxies. 
}
\label{fig:EWprof_comb}
\end{figure*}

To classify the galaxies in their corresponding EL classes, we use set of figures and diagrams for each galaxy as discussed in Sec. \ref{S:2DELC}. One important element is the W$_{\rm H\alpha}$ radial profile, which provides a detailed information about the distribution of the retired and the star-forming regions of the galaxies.

In Fig.~\ref{fig:EWprof_matrix}, we show the characteristic (stacked) W$_{\rm H\alpha}$ radial profiles of all galaxies in each EL class. The profiles are obtained through a radial map positioned in the centre of the galaxy and projected on the plane of the galaxy, assuming the position angle and inclination reported in Table B of \citetalias{Kalinova2017b}. The radial maps are then normalised by the galaxy effective radius (see Table \ref{tab:long}). The profiles are constructed by considering all the W$_{\rm H\alpha}$ map spaxels in bins of 0.25 R$_e$ for each galaxy in a given EL class (dotted vertical lines Fig.~\ref{fig:EWprof_matrix} indicate the bin step). The coloured,  full lines show the median of the radial distributions, the coloured, dashed lines indicate the $25^{\rm th}$ and $75^{\rm th}$ percentiles of the radial distributions.
Similarly to Fig. \ref{fig:2DELC},  the rows of Fig~\ref{fig:EWprof_matrix} correspond to the quenching stages of the galaxies, and the columns correspond to their nuclear activity groups. The nuclear activity leaves a clear trace within roughly 0.2 $R_e$ of the profiles:  W$_{\rm H\alpha}$ $>3$ {\AA} for weak AGN (middle column),   W$_{\rm H\alpha} >6$ {\AA} for strong AGN (right column),  while this value is not fixed for the non-active galaxies (left  column). 

Further, we compare the median profiles of the galaxy in each quenching stage for the three different nuclear activity groups in Fig.~\ref{fig:EWprof_comb}. 
The details of those profiles vary significantly between the classes (especially within 0.5 $R_e$), however the average value of W$_{\rm H\alpha}$ generally increases from the centre of the galaxy to the outskirts. This might indicate that the star formation quenching in the galaxies largely proceeds ``inside-out'', which has been noticed for CALIFA galaxies elsewhere (e.g. \citealt{Gonzalez-Delgado2016}, \citealt{Belfiore2017a}, \citealt{Lin2017}, \citealt{Ellison2018}, \citealt{Sanchez2018}), however patterns can be more complicate as shown by the mixed and retired classes. 

The star-forming galaxy group shows the highest radial values of W$_{\rm H\alpha}$, well above the 6 {\AA} threshold (see also panels A, B, and C in Fig. \ref{fig:EWprof_matrix}), where the radial median of W$_{\rm H\alpha}$ is larger than 13 {\AA} at each radii). This is particularly clear for the non-active group in the left panel of Fig.~\ref{fig:EWprof_comb}, which contains the largest statistical sample of galaxies.

In particular considering the nuclear activity, the median profile of the sAGN galaxies is between 3 and 6 {\AA}, which is below the one from the nonA galaxies across the radial extend of the profile. The wAGN profile is roughly similar to the nonA profile for the SF quenching stage, except within 0.5\,$R_e$ where they differ with $\sim10$ {\AA}. Nevertheless, for the other quenching stages the difference between the profiles from the three nuclear activity groups are less evident.

Similarly high values (W$_{\rm H\alpha}$ $\sim$ 10 {\AA}) are reached by the median radial profiles of centrally-quiescent and quiescent-nuclear-ring classes beyond one effective radius. Within 1~$R_e$, instead, both profiles show a dip below W$_{\rm H\alpha}=3$ {\AA} that characterises quenched regions (see also panels D and G in Fig. \ref{fig:EWprof_matrix}. Nevertheless, the central radial bin (where $R=0.25~R_e$) of the quiescent-nuclear-ring class shows values again above the threshold that defined star formation dominance (6 {\AA}).

It is interesting to note the case of the mixed profile, which is constantly between 3 {\AA} $ <$W$_{\rm H\alpha} \leq$ 6  {\AA}, except in the central bins where the radial median of W$_{\rm H\alpha}$ is below 3 {\AA} (see also panel F of Fig. \ref{fig:EWprof_matrix}). This might be an indication that also for this class, that generally presents a complex star formation pattern, the quenching begins from the very centre. Further, in Fig. \ref{fig:EWprof_comb},  the mixed$-$wAGN profile in the regions of $R>2~R_e$  goes  above the W$_{\rm H\alpha}=6 $ {\AA} demarcation line (see also Fig.~\ref{fig:EWprof_matrix}, panel I). This profile is generated by a single galaxy and may not be representative of the mixed-wAGN category. 

Nearly-retired and fully-retired galaxy groups show W$_{\rm H\alpha}$ almost constantly below the 3 {\AA} threshold that marks ionisation by post AGB stars dominance. Both profiles increase in values from the centre to the outskirt, but the  nearly-retired profile is always above the fully-retired median profile. In addition the nearly-retired galaxy group profile reaches value of W$_{\rm H\alpha}>3$ {\AA}, where $R>3~R_e$, which might indicate some weak SF activity in this region. The profiles from the AGN host galaxies in the nearly-retired stage are basically equivalent to the non-active counterparts.

\section{Global properties of the galaxies at different quenching stages}
\label{S:ELCprop}
To understand the typical properties of each EL class, we investigate some basic parameters of the galaxies. Several galactic properties show, on average, monotonous behaviours through the EL classes. This indicates that \emph{QuestNA} is a valid classification method for galaxies, as for a given class corresponds well defined set of galactic properties.

\begin{figure}
\centering
\includegraphics[width=0.43\textwidth]{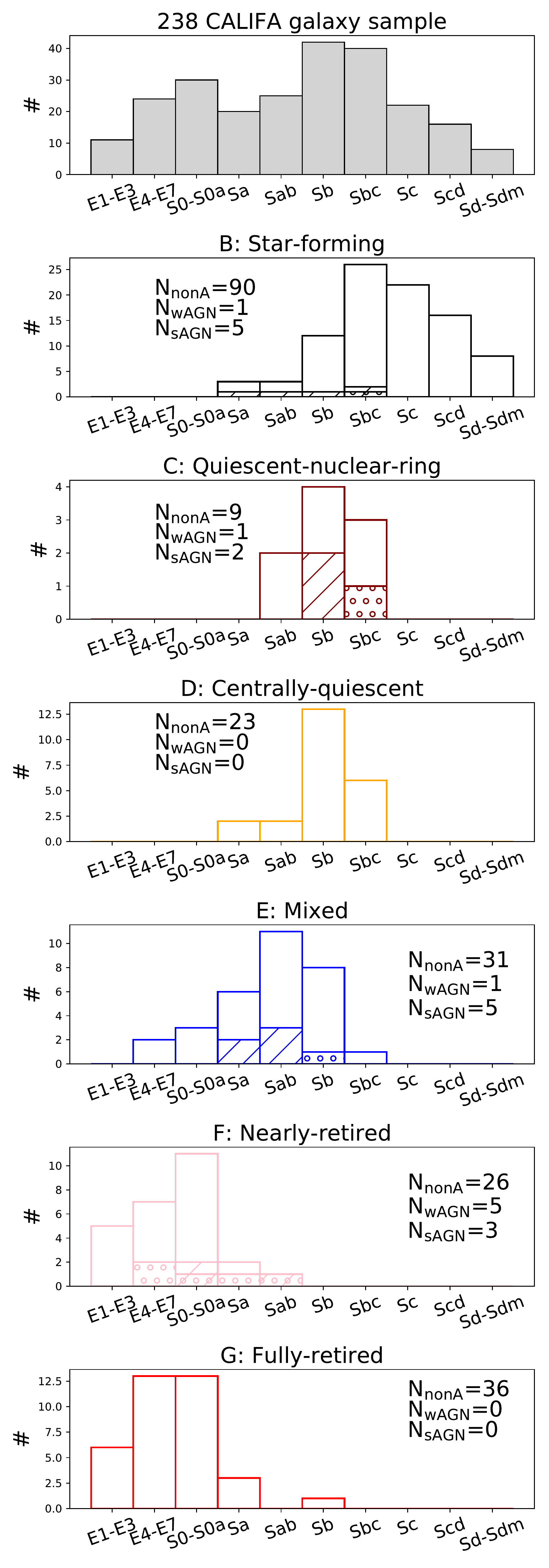}
\caption{Distributions of the active and non-active galaxies in each quenching stage through Hubble type. The continuous line outline the distribution of the non-active galaxies, while the circles and the slashes correspond to the wAGN and sAGN distributions, respectively. $N$ gives the number of the galaxies for each nuclear activity group (see Sec. \ref{SS:morph}). }
\label{fig:Histos_EL}
\end{figure}

\begin{figure}[h]
\centering
\includegraphics[width=0.48\textwidth]{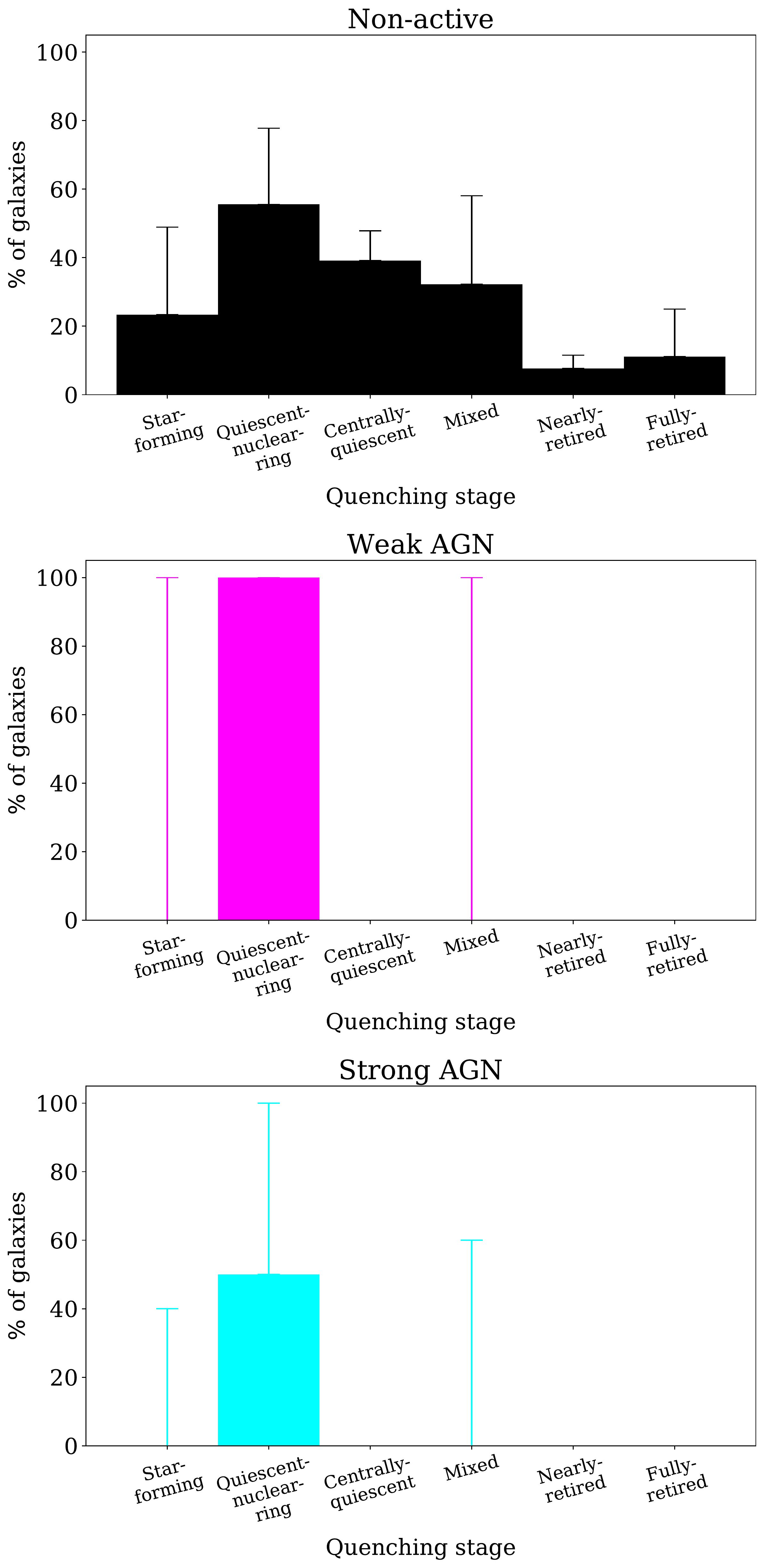}
\caption{Distributions of the barred (B) galaxies through active and non-active galaxies for each quenching stage (QS) of our sample. The upper error-bar for each QS indicates the fraction of galaxies with the unsure bar (AB). Quiescent-nuclear-ring is represented by the highest number of barred galaxies, followed by centrally-quiescent and mixed groups (see Sec \ref{SS:morph}).}
\label{fig:Bar_histo}
\end{figure}

\begin{figure*}
\centering
\includegraphics[width=0.95\textwidth]{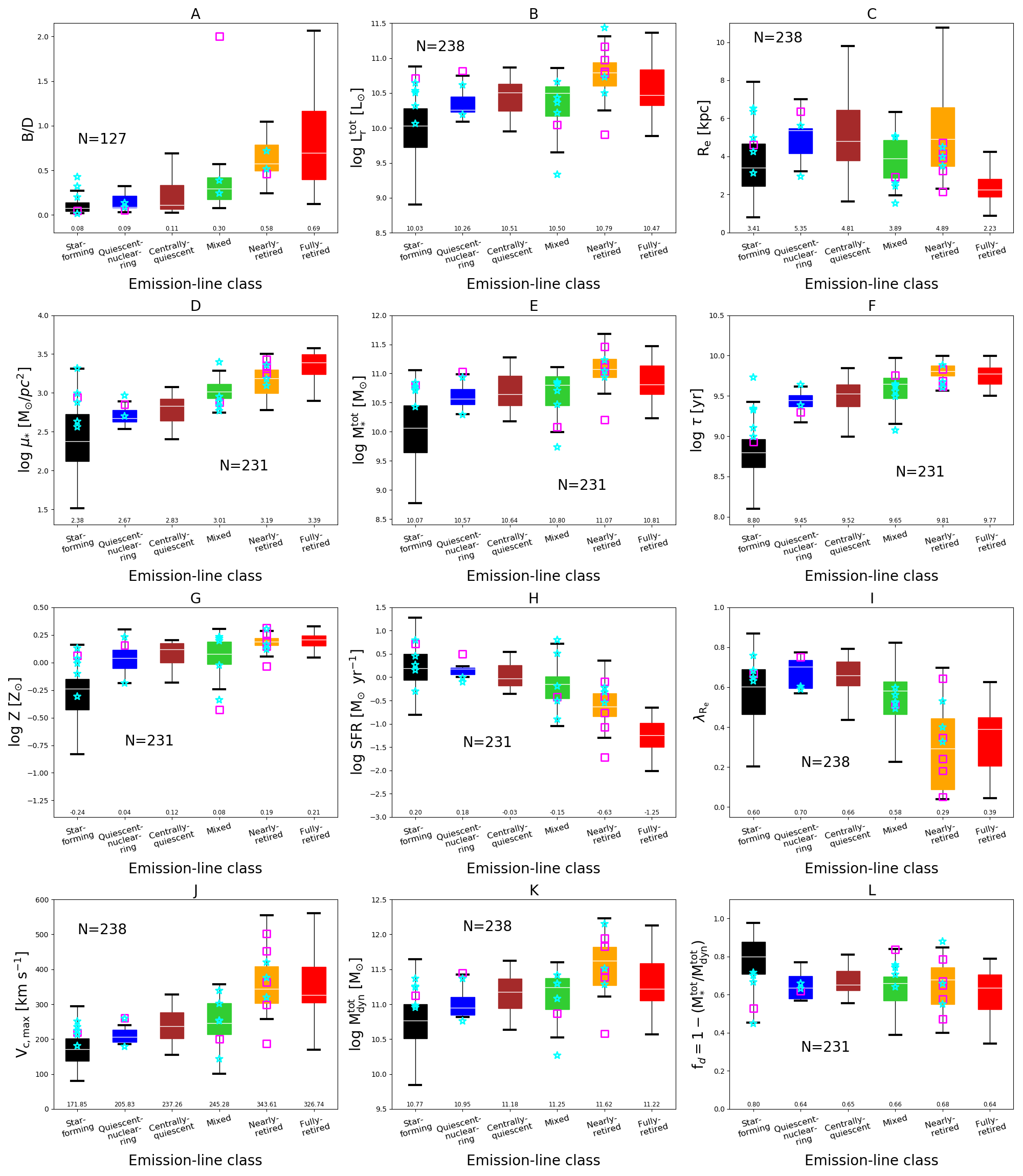}
\caption{Box-and-Whisker diagrams of basic galaxy properties through the EL classification (see Sec. \ref{S:ELCprop}). The colour diagrams represent the properties of the non-active galaxies as follows: black $-$ star-forming; dark blue $-$  quiescent-nuclear-ring; brown $-$  centrally-quiescent;  green $-$  mixed; orange $-$ nearly-retired; and red $-$  fully-retired. The active galaxies are over-plotted with empty symbols onto the diagrams for a given quenching stage (cyan star for sAGN and magenta square for wAGN). The boxes represent the data between the lower and the upper quartiles of the distribution with median values (\citealt{Tukey1977}), given by the white lines and the numbers. The extend of the data outside from the quartiles (whiskers) are indicated by the vertical black lines.  The label ``$\emph{N}$'' in each panel gives the total number of the galaxies, participating in the statistics. These diagrams represent distributions of various galactic properties in the different emission-line classes: bulge-over-disk flux ratio (B/D, panel A);   total luminosity in $r-$band ($L_{\rm r}^{\rm tot}$, panel B); effective radius ($R_e$, panel C); stellar mass surface density ($\mu_*$, panel D); stellar mass ($M_*$, panel E); stellar age ($\tau$, panel F); stellar metallicity (Z, panel G); star formation rate (SFR, panel H); specific angular momentum ($\lambda_{\rm R_e}$, panel I); peak circular velocity ($V_{\rm c, max}$, panel J); total dynamical mass ($M_{\rm dyn}^{\rm tot}$, panel K); mass discrepancy factor ($f_{\rm d}$, panel L). The SF$-$sAGN galaxies shows the most discrepant behaviour with respect to their non-active counterparts.
The median values of the galaxy properties from the Box-and-Whisker diagrams are given in Table \ref{tab:Box-medians}. See texts in Sec. \ref{S:data} and \ref{S:ELCprop} for further details. }
\label{fig:Box}
\end{figure*}

\subsection{Morphology}
\label{SS:morph}
The morphology of similar emission-line class galaxies has been broadly explored in the literature (e.g., \citealt{Schawinski2007}, \citealt{Sanchez2018}, \citealt{Sanchez2019}, \citealt{Lacerda2020}), and it is an important parameter that gives a direct information about the structural characteristics of the galaxies in the various categories. In Fig. \ref{fig:Histos_EL}, we show the distributions of our EL classes through the Hubble sequence using the CALIFA classification of \cite{Walcher2014}. In general, the morphology changes from late- to early-type moving from the star-forming to the retired classes.

Examining the non-active galaxies (continuous lines is Fig. \ref{fig:Histos_EL}), the star-forming class are mainly late-type spirals (Sb--Sdm), while quiescent-nuclear-ring and centrally-quiescent systems are represented by early-type spirals (Sab--Sbc). Similarly to the  centrally-quiescent and quiescent-nuclear-ring objects, the mixed class possess early-type spiral morphologies (Sa--Sb), but it also includes few cases of lenticulars (S0--S0a) and ellipticals (E4--E7). Furthermore, the nearly-retired and fully-retired classes of the non-active galaxies are mainly presented by lenticular and elliptical galaxies.

In total, sAGN class (slashes in Fig. \ref{fig:Histos_EL}) is mostly represented by early-type spiral (Sa--Sbc) and lenticular (S0--S0a) galaxies. Similarly, wAGN class (circles in Fig. \ref{fig:Histos_EL}) spans over the early-type spiral (Sa--Sbc) and lenticular (S0-S0a) systems, but it also includes older galaxies as the ellipticals (E4--E7).
The comparison between the distributions of the active and non-active galaxies shows a mismatch for the star-forming and nearly-retired QSs. Panel B of Fig. \ref{fig:Histos_EL} of the star-forming QS shows that the sAGNs are represented by early-type spiral galaxies (Sa--Sbc), which differs from the main peak of the non-active galaxies' distribution, located towards the late-type spirals. 
Furthermore, Panel F of Fig. \ref{fig:Histos_EL} of the nearly-retired QS indicates that the wAGNs span over the early-type galaxies and early-type spirals, where the latter is an outlier region for the non-active galaxies' distribution,  peaking at the lenticular galaxies.

We also make a structural comparison between the active and non-active group, taking into account the presence of a sure (B) or unsure (AB) bar in each quenching group based on CALIFA bar classification of \cite{Walcher2014}. 
Fig. \ref{fig:Bar_histo} shows that the highest number of barred galaxies belong to the  quiescent-nuclear-ring QS, followed by the centrally-quiescent and mixed groups. The upper error-bar of the histograms represents the maximal number of bars in a group, which include both bars (B) and unsure bar (AB) cases of the galaxies (see also Table \ref{tab:long}).
The information from the morphological and bar classifications of  \cite{Walcher2014} for our sample, suggest that the secular evolutionary processes (which involve the internal evolutionary phenomena induced by dynamical features such as spiral arms and bars) are much more prominent in the quiescent-nuclear-ring, centrally-quiescent and mixed classes than in the rest of the classes.

\subsection{Photometric, stellar, kinematic, and dynamical properties}
\label{SS:phot}
Besides the galaxy morphology, our classification correlates with a large set of galactic properties as shown in Fig.~\ref{fig:Box}. This proves that knowing the quenching stage of a galaxy is sufficient to access the average value of a variety of parameters. The bulge-to-disk ratios (B/D, panel A); the stellar mass surface density ($\mu_*$, panel D); the stellar age ($\tau$, panel F); the stellar metallicity ($Z$, panel G); and the peak circular velocity ($V_{\rm c,max}$, panel J)  all increase moving from the star-forming to the fully-retired groups for the non-active galaxies. On the opposite, the SFR sharply decreases across the quenching stages (panel H). 

For other properties the trends are less obvious. Regarding the total luminosity (L$_{r}^{\rm tot}$, panel B), the brightest galaxies, on average, are observed within the nearly-retired class, and not within the fully-retired ones, while star-forming class are the faintest objects. A similar trend is shown by the total stellar and dynamical mass ($M_*$, panel E; and $M_{\rm dyn}^{\rm tot}$, panel K; respectively), while the opposite behaviour is observed for the angular momentum ($\lambda_{\rm R_e}$, panel I). The size of the galaxies, inferred from the median effective radius, $R_e$, shows a complicated behaviour across the stages (panel C).  The largest-size galaxies are mainly quiescent-nuclear-ring galaxies (R$_e\sim$5.35 kpc) and nearly-retired (R$_e\sim$4.89 kpc), followed by the centrally-quiescent (R$_e\sim$4.81 kpc). Fully-retired galaxies have the smallest  R$_e\sim$2.23 kpc in comparison to the rest of the classes, which might indicate that most of the luminosity is generated by their large spheroidal components (considering their large B/D ratio). 

The mass discrepancy factor $f_d=1-(\mathrm{M^{tot}_*/M^{tot}_{dyn}})$ (panel L) displays a rather flat behaviour. We find that $f_d$ is largest for star-forming ($f_d=$80 \%) in comparison to the rest of the quenching classes, where $f_d$ varies between 64 and 68 \%. The large value of the $f_d$ through EL classes could be due to non-universal nature of the galaxy IMF (e.g., \citealt{Hoversten2008}, \citealt{Cappellari2012b}, \citealt{Lyubenova2016}), radial variation of stellar mass-to-light ratio $\gamma_{*}$ (e.g., \citealt{Garcia-Benito2019}) or the presence of dark matter mass (e.g., \citealt{Das2020}). 
The variation of the mass discrepancy could be also explained with an uncounted gas mass in the galaxies. It is expected to be significant for late-type systems (e.g. \citealt{McGaugh2000}). 
In forthcoming papers, we will infer the total gas mass fractions of our sample using a dust-to-gas calibration (see \citealt{Barrera-Ballesteros2018}) or 
 CO data (from \citealt{Bolatto2017}, \citealt{Colombo2020}) and HI  data (Kalinova et al., in prep) to calculate the dark matter content for each galaxy. 

The median values of the galaxy properties from the Box-and-Whisker diagrams are listed in Table \ref{tab:Box-medians}.

The behaviour of the AGN-host galaxies at a given quenching stage is also reported in Fig.~\ref{fig:Box}. The sAGN galaxies within the star-forming class are the objects that shows the most discrepant behaviour with respect to their non-active counterparts. Those galaxies are mostly located in the upper end of the distributions of all parameters, except for the SFR (panel H), where sAGN-galaxies show global SFRs similar to the non-active objects in the same class. However,  star-forming-sAGN appear between the objects of the star-forming class with the lowest $f_d$. AGN-host galaxies that belong to the mixed and nearly-retired categories seem to show low values of $R_e$.

\section{Discussion}
\label{S:disc}
\begin{figure*}
\centering
\includegraphics[width=0.95\textwidth]{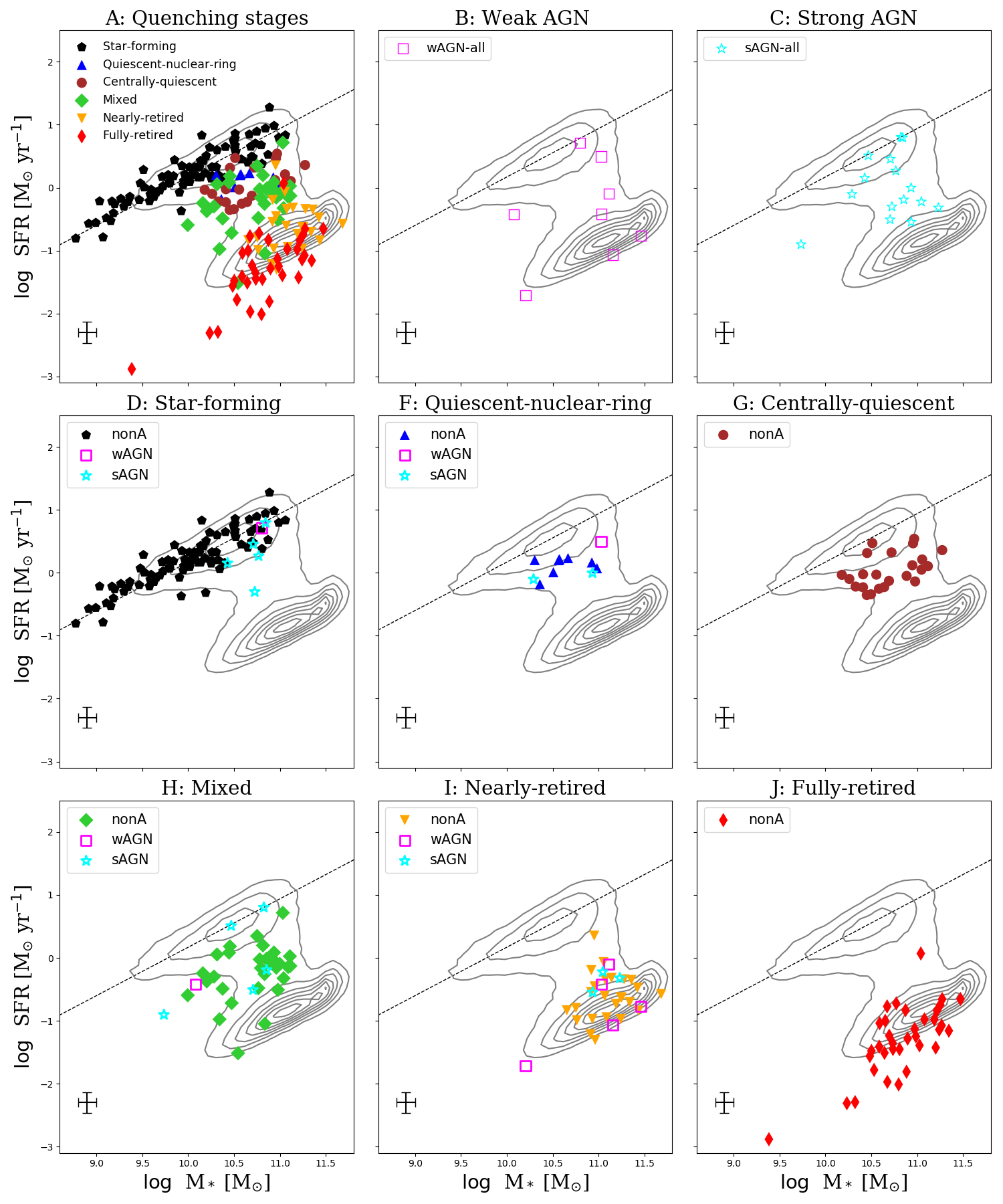}
\caption{The SFR-$M_*$ diagram for our sample of CALIFA galaxies, considering their quenching stages and nuclear activities (see Sec. \ref{S:SFRM}).  All active and non-active galaxies through quenching stage are presented in panel A. Panel B and C display the galaxies that host a weak or strong AGN (respectively), independently of their quenching stage. Panels D$-$J show the galaxies in a given quenching stage, where the filled symbols represent the non-active galaxies, while the empty symbols correspond to the active galaxies: wAGN (magenta squares) or sAGN (cyan stars). The dashed line indicate the main sequence (MS) of star formation using the fit of \citet{Elbaz2007}. The grey contours represents the SDSS DR7 population for reference from MPA-JHU catalogue (\citealt{Kauffmann2003,Brinchmann2004,Salim2007}). The error bars in the panels represent the typical uncertainties of the stellar masses and the SFRs, adopted from \citealt{CidFernandes2014} and \citealt{Gonzalez-Delgado2017}, respectively. The degree of the star-formation quenching increases along the present EL-pattern sequence from star-forming to fully-retired. For a given quenching stage, the active galaxies occupy the same domain on the SFR-M$_*$ diagram as the non-active ones. }
\label{fig:MS}
\end{figure*}

\begin{figure*}
\centering
\includegraphics[width=0.8\textwidth]{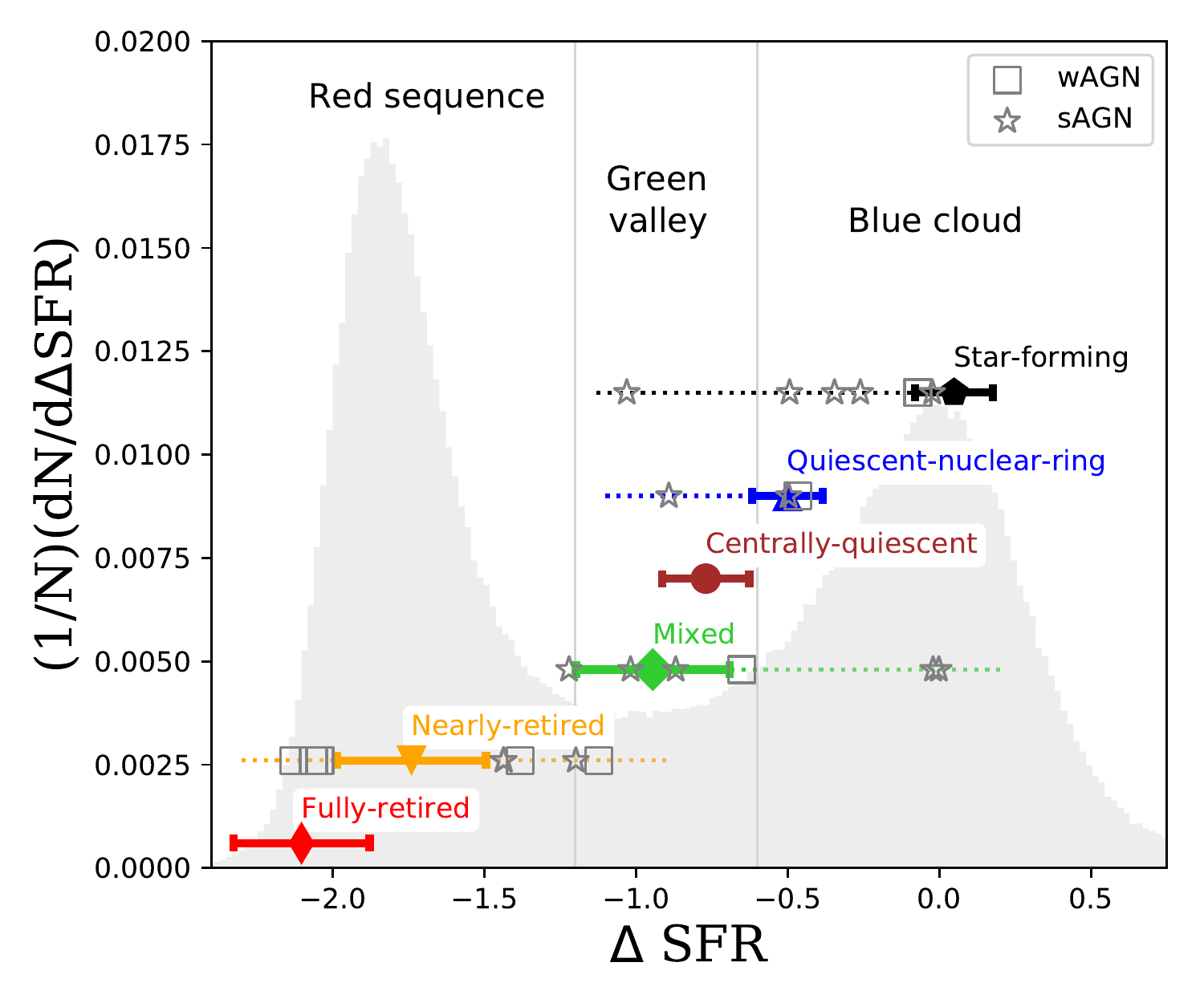}
\caption{Distribution of $\Delta$ SFR for SDSS DR7 galaxies at z$\sim$0 (shaded regions; MPA-JHU catalogue, \citealt{Kauffmann2003,Brinchmann2004,Salim2007}) and for our sample of galaxies. The SDSS y-axis gives the number of the galaxies  per given $\Delta$SFR bin (i.e., dN/d$\Delta$SFR), normalised to the total number N of the SDSS galaxies, while the y-axis of our sample is arbitrary and chosen here only for representative purposes (due to the smaller number of our galaxies in contrast to the number of the SDSS galaxies). The vertical grey lines indicate the Green valley region (adopted values from \citealt{Bluck2016}). 
The colour symbols indicate the median $\Delta$ SFR distribution for the non-active galaxies of our sample in each quenching stage, while the colour bar indicate the corresponding median absolute deviation. The active galaxies are presented by the grey empty symbols (stars for sAGN and square for wAGN). The horizontal dotted colour lines are plotted to guide the eye for the position of the active galaxies in comparison to the non-active ones for a given quenching stage. 
Quenching stage galaxies are well divided through $\Delta$ SFR, suggesting that the quenching stage represent a certain evolutionary phase from the life-cycle of the galaxy  (see Sec. \ref{S:SFRM}).
}
\label{fig:Histo}
\end{figure*}

\subsection{Imprints of galaxy evolution in the EL classes}
\label{S:SFRM}
To understand whether our proposed categories are linked to galaxy evolution, we superimpose the EL classes on the SFR-M$_*$ diagram (Fig.\ref{fig:MS}). Galaxies in the star-forming class almost fully populate the star formation main sequence (MS; Fig.\ref{fig:MS}, panel B). However, the AGN-host galaxies at this quenching stage show mostly a strong AGN activity and they are located in the upper end of the main sequence, occupied by high-mass galaxies.  Quiescent-nuclear-ring and centrally-quiescent galaxies appear to populate the so-called ``Green valley'' of star formation, which is generated by objects that are switching off their star formation in the central region (Fig.\ref{fig:MS}, panel C and D, respectively). Galaxies with a undefined distribution of the ionisation  (mixed class) spread from the MS to the red sequence, but they appear mostly located in the ``Green valley'' as well (Fig.\ref{fig:MS}, panel E). Strong AGN-host galaxies appear well distributed across the full high-mass region of the SFR-$M_*$ diagrams.  Nearly-retired and fully-retired galaxies are in tight constrain along the red sequence. In particular, nearly- and fully-retired galaxies dominate the high-mass and low-mass end of the Red sequence, respectively. 
It is interesting to note that the  nearly-retired galaxies are the most massive systems from our sample, which are also the most dispersion-dominated galaxies (i.e., they have the lowest $\lambda_{R_e}$, panel I of Fig.\ref{fig:Box}) with the largest size stellar disks among the other classes (panel A and C of Fig.\ref{fig:Box}). They are old stellar population galaxies, but with still little SF within the 2 R$_e$. This gives a clue that maybe the fully-retired class are the progenitor of the nearly-retired  class, where SF have been re-ignited  in the galaxies by merger events or environmental processes in earlier epochs, or the two classes might have had independent evolutionary paths since the beginning of their formation (or it is a combination of both mechanisms). To understand more about the origin of the nearly- and fully-retired galaxies, in our next article we will link the QS of the galaxies and  their dynamical properties  (Kalinova et al., in prep).

Further, AGN galaxies at the  nearly-retired stage appear to be mostly in the ``weak'' category. This might indicate that the AGN activity reduce with the evolution of the galaxies and the progression of the quenching, possibly due to the progressive diminishing of the gas supply.

Panels B and C collect the position of all wAGN and sAGN galaxies, independently of their quenching stage. The location of the AGN host galaxies in the transition region between the Blue cloud and the Red sequence has been reported in other works too (e.g., \citealt{Schawinski2014,Sanchez2018,Cano-Diaz2019,Lacerda2020}).
The wAGN galaxies are largely dispersed between the Green valley and the Red sequence, and mostly occupying either the low-mass or high-mass end of the SFR-M$_*$ diagram. Further, they also show bi-modality in their properties (see also \citealt{Heckman2014}).

A quicker way to investigate the properties of the galaxies in respect to their position on the SFR-M$_*$ diagram and connect to their stage of evolution (e.g., \citealt{Ellison2018},  \citealt{Thorp2019}, \citealt{Bluck2020}) is provided by Fig. \ref{fig:Histo}, which summarises the distribution of $\Delta$ SFR, where $\Delta$ SFR $\equiv \log$ SFR - $\log$ MS. $\Delta$ SFR is the logarithmic distance from the star-forming main sequence (MS) of \citet{Elbaz2007} using the method described in \citet{Bluck2016}. In the figure, the normalised histogram represents local galaxies from SDSS DR7 with a given $\Delta$ SFR. In essence, the SDSS histogram in grey highlights the fact that galaxies at $z\sim0$ are mostly located in the Red sequence, followed by the Blue cloud, and the Green valley. In the figure, the quenching stages of non-active galaxies are indicated with the median and the median absolute deviation of $\Delta$ SFR of the given sub-samples.  QS galaxies appear well divided through $\Delta$ SFR, indicating how a given quenching stage represents a relatively narrow range of certain physical parameters, validating our classification as a useful tool to trace galaxy evolution through the star formation and nuclear activity. Only the median absolute deviation bars of mixed and centrally-quiescent classes overlap in the Green valley region.

Furthermore, in Fig. \ref{fig:Histo} for a given QS, we explore how the $\Delta$SFR values of the active galaxies (grey star for sAGN and grey square for wAGN) are comparable with the $\Delta$SFR median value (colour symbols) and median absolute deviation (colour bars) of the non-active systems. Overall, star-forming sAGN galaxies show the biggest mismatch with the distribution of their non-active counter-partners. These type of sAGN galaxies are all shifted towards lower $\Delta$SFR values (i.e., they all deviate from the Main Sequence), with respect to the star-forming non-active objects. This suggests that the suppression of the star-formation have already started in these galaxies and it might be related to the presence of the AGN in their centres through feedback processes (e.g., \citealt{Frigo2019}, \citealt{Kaufmann2000}).

\subsection{Inside-out quenching and the role of AGN}
Through our EL classification, we observe that the galaxies on the way to quenching always begin to switch off their star formation from the centre (see Fig. \ref{fig:EWprof_comb}). 
The Quiescent-nuclear-Ring galaxies are an exception (since they have some instance of star formation  below 0.2 R$_e$) but also a rare example, and they are mostly quenched within 0.3$-$0.5 R$_e$. In other words, we rarely observed a galaxy that is largely star forming in the centre and fully quenched in the outskirts. Usually, this is due to the presence of a close companion or interaction between galaxies (which have been purposely excluded by the sample analysed here). Outside-in SF quenching may also be induced by ram-pressure stripping in galaxy cluster's environment (e.g., \citealt{Bekki2009}, \citealt{Vulcani2020}).  

Our results are consistent with an ``inside-out'' quenching scenario (e.g., \citealt{Gonzalez-Delgado2016}, \citealt{Catalan-Torrecilla2017},  \citealt{Belfiore2018}, \citealt{Sanchez2018}, \citealt{Lin2019}, \citealt{Lacerda2020}), which predicts that the star formation quenching starts from the centre of the galaxies due to the presence and the continuous growth of the bulge (e.g., \citealt{Martig2009}). 
Galaxies with larger central mass concentration are expected to host more massive black holes (BH) from the M$_{bulge}$-M$_{BH}$ relation 
(e.g.,  \citealt{Marconi2003}; \citealt{Haring2004}; \citealt{McConnell2013}), and experience more significant energetic feedback. 
Thus, the growth of the central spheroid could explain the precedence of the "inside-out" over the "outside-in" quenching scenario for the evolution of our galaxies (e.g., \citealt{Gonzalez-Delgado2016}, 
\citealt{Ellison2018}, \citealt{Lin2019}, \citealt{Bluck2019}).

The proposed EL classification is mostly based on the morphological distribution of the value of the W$_{\rm H\alpha}$. This approach presents a unique perspective that allows us to define some EL classes as specific ``quenching stages''. Nevertheless, we also include the BPT diagrams to search for signatures of AGN ionisation in the galaxies. 
In particular, while sAGN are exclusively observed in the Green valley of our sample (see also \citealt{Kauffmann2003}, \citealt{Salim2007}, \citealt{Sanchez2018}, \citealt{Lacerda2020}), wAGN host galaxies 
are spread over the Green valley and the Red sequence. A similar bi-modality of the AGNs is also noted by  \cite{Heckman2014}. 

The role of the AGN in the quenching of the galaxies is complex and not yet well understood. AGN can enhance or suppress SF in the galaxies (e.g., \citealt{Wagner2016}, \citealt{Husemann2018}, \citealt{Cresci2018}). For instance,  gas outflow from the AGN can trigger formation of new stars in dense cold gas environments (the so called positive feedback; \citealt{Silk2013}, \citealt{Zubovas2013}, \citealt{Mahoro2017}). On the other hand, the suppression of the SF could happen via heating of the cold gas or mechanically removing of the gas (the so called negative feedback; \citealt{Silk1998}, \citealt{Nesvadba2010}, 
\citealt{McNamara2016}).

The negative feedback is well accepted in models to regulate massive galaxy growth (e.g., \citealt{Kaufmann2000}, \citealt{Schaye2015}).
For example, \cite{Frigo2019} investigate the impact of the AGN feedback on the formation of massive galaxies through performance of cosmological simulations and find evidence that the AGN feedback plays an essential role in the formation of quiescent, slow-rotating, triaxial and non-discy early-type galaxies.

We found only a small percentage of AGN-host galaxies in our sample. Similar small percentages has been observed elsewhere, and they have been interpreted as evidence that the star formation quenching in the nearby universe cannot be due to AGN action alone (e.g., 
\citealt{Saintonge2012}, \citealt{Kirkpatrick2014}, \citealt{Saintonge2017}). Some studies report that AGN feedback acts locally  and does not influence the global SFR budget of the galaxy (e.g., \citealt{Shin2019}). However,  these conclusions could be misleading, since the AGN may no longer be luminous, when the effect from the AGN becomes observable (as discussed in \citealt{Harrison2017}). 

Surprisingly, the active and non-active galaxies at a given phase of their evolution show similar quenching patterns, or in other words, active galaxies do not show peculiar quenching patterns that are not amenable to one of the main six ``quenching stages'' that we have defined. Additionally, the analysis of active and non-active galaxies at the same quenching stages, allowed us to observe that there are not significant differences in the properties of the objects with different nuclear activity. An exception is observed for the SF stage, where all AGN (mainly strong AGN) in this group are hosted by high mass and older galaxies, quite close to the star formation ``green valley''. It seems that galaxies do not develop an observable AGN until given values of several properties are reached (e.g. so-called "Golden mass", \citealt{Dekel2019}).  

Future exploration of the gas properties of the EL galaxies will shed more light on the relation between the role of the AGN feedback and the SF quenching of the galaxies.

\section{Summary}
\label{S:summary}
We explore the ionised gas properties of 238 (E1$-$Sdm) CALIFA galaxies through performing a 2D emission-line classification (\emph{QuestNA}), which is based on the spatially resolved W$_{\rm{H\alpha}}$ maps and BPT diagrams ([NI], [SII], [OI]). We define six quenching stages: galaxies dominated by recent star-formation; centrally-quiescent galaxies; systems possessing quiescent-nuclear-ring-like structure; galaxies with mixed ionised processes; nearly-retired galaxy that posses only a few star forming regions, or systems that are completely quiescent (fully-retired within 2 R$_e$). We further divide the galaxy according to their nuclear activity: non-active (nonA) galaxies, host of a weak (wAGN) or strong AGN (sAGN). The degree of the star-formation quenching increases along the present EL-pattern sequence from  star-forming to fully-retired. 

The EL classes occupy certain domains in the SFR-M$_*$ diagram, where star-forming and quiescent-nuclear-ring classes are located within the Blue cloud; centrally-quiescent and mixed are transient galaxies in the Green valley; nearly- and fully-retired are members of the Red sequence.

The active galaxies do not show significantly different properties from the corresponding non-active counterparts for a given quenching stage, except for the SF$-$sAGN class. Overall, the sAGN host galaxies are mainly located in the Blue cloud or the Green valley. On the other hand, the wAGN galaxies are largely dispersed between the Green valley and the Red sequence, and mostly occupying either the low-mass or high-mass end of the SFR-M$_*$ diagram.

Morphological analysis of our sample indicates that secular evolutionary features (e.g., bar, spiral arms) dominate  the structure of quiescent-nuclear-ring, centrally-quiescent and mixed classes in comparison to the other EL classes. The W$_{\rm{H\alpha}}$ profiles of the EL galaxies indicate an  "inside-out" SF quenching scenario and emphasise the importance of the galaxy bulge/central spheroid (see Fig. \ref{fig:EWprof_comb}). 

In forthcoming papers, we will continue the exploration of the CALIFA galaxies grouped in EL classes through various other properties such as dynamics, velocity profiles, gas content and  environment, in order to complete the proposed classification and further explore the quenching mechanism of the active and non-active galaxies.

\section*{Acknowledgements}
\label{S:acknowledgements}
We thank the anonymous referee, whose suggestions helped us to improve the quality and presentation of this paper.
DC acknowledges support by the \emph{Deut\-sche For\-schungs\-ge\-mein\-schaft, DFG\/} project number SFB956A.
SFS acknowledge Conacyt projects CB-285080 and FC-2016-01 -1916, and PAPIIT-DGAPA IN100519.
RGB and RGD acknowledges financial support from the State Agency for Research of the Spanish MCIU through 
the ``Center of Excellence Severo Ochoa'' award to the Instituto de Astrof\'isica de Andaluc\'ia (SEV-2017-0709), and  AYA2016-77846-P.
ER acknowledges the support of the Natural Sciences and Engineering Research Council of Canada (NSERC), funding reference number RGPIN-2017-03987.
 
In this study, we made use of the data of the first legacy survey, 
the Calar Alto Legacy Integral Field Area (CALIFA) survey, 
based on observations made at the Centro Astron\'omico
Hispano Alem\'an (CAHA) at Calar Alto, operated jointly by the Max Planck-Institut
f\"ur Astronomie and the Instituto de Astrof\'isica de Andaluc\'ia
(CSIC).

This research has made use of the NASA/IPAC Extragalactic Database (NED), which is operated by the Jet Propulsion Laboratory, California Institute of Technology, under contract with the National Aeronautics and Space Administration.

Funding for the Sloan Digital Sky Survey IV has been provided by
the Alfred P. Sloan Foundation, the U.S. Department of Energy Office of
Science, and the Participating Institutions. SDSS-IV acknowledges
support and resources from the Center for High-Performance Computing at
the University of Utah. The SDSS web site is www.sdss.org.
SDSS-IV is managed by the Astrophysical Research Consortium for the 
Participating Institutions of the SDSS Collaboration including the 
Brazilian Participation Group, the Carnegie Institution for Science, 
Carnegie Mellon University, the Chilean Participation Group, the French Participation Group, Harvard-Smithsonian Center for Astrophysics, 
Instituto de Astrof\'isica de Canarias, The Johns Hopkins University, 
Kavli Institute for the Physics and Mathematics of the Universe (IPMU) / 
University of Tokyo, Lawrence Berkeley National Laboratory, 
Leibniz Institut f\"ur Astrophysik Potsdam (AIP),  
Max-Planck-Institut f\"ur Astronomie (MPIA Heidelberg), 
Max-Planck-Institut f\"ur Astrophysik (MPA Garching), 
Max-Planck-Institut f\"ur Extraterrestrische Physik (MPE), 
National Astronomical Observatory of China, New Mexico State University, 
New York University, University of Notre Dame, 
Observat\'ario Nacional / MCTI, The Ohio State University, 
Pennsylvania State University, Shanghai Astronomical Observatory, 
United Kingdom Participation Group,
Universidad Nacional Aut\'onoma de M\'exico, University of Arizona, 
University of Colorado Boulder, University of Oxford, University of Portsmouth, 
University of Utah, University of Virginia, University of Washington, University of Wisconsin, 
Vanderbilt University, and Yale University.\\
This research made use of Astropy (\url{http://www.astropy.org}), a community-developed core Python package for Astronomy (Astropy Collaboration, 2013).

\footnotesize{
\bibliographystyle{aa}
\bibliography{SFQS.bib}
}

 \appendix
 \label{A:appendix}

\section{Diagnostic maps and diagrams for all 238 CALIFA galaxies (Online material)}
\label{A:EL_examples}

We present the resolved $W_{H\alpha}$ maps, $W_{H\alpha}$ profiles, BPT-maps and BPT-diagrams for the full sample of 238 CALIFA galaxies.\\

 (\emph{The online material is available in the published version of the paper and upon request from the authors}).

\section{Tables with properties of the 238 CALIFA galaxies}
\label{A:table}
In Table \ref{tab:long} and  \ref{tab:Box-medians}  , we list the main properties of the 238 (E1$-$Sdm) CALIFA galaxies, presented in this work.

\onecolumn

\begin{center}
\begin{longtable}{|l|l|l|l|l|l|l|l|l|l|l|l|}
\caption{Properties of the 238 (E1-Sdm) CALIFA galaxies.
Columns list: 
(1) Galaxy identifier (NED);
(2) Galactocentric galaxy distance in Mpc (NED);
(3) Effective radius of the galaxy in arcseconds, measured from SDSS images via growth curve analysis using elliptical apertures (Sanchez et al., in prep);
(4) Stellar mass, adopting Chabrier IMF in M$_{\odot}$ (Sec. \ref{SS:properties_pop});
(5) Star Formation Rate of the galaxies in M$_{\odot}$ yr$^{-1}$ (Sec. \ref{SS:properties_pop});
(6) Star Formation Rate distance, measured from the Main Sequence in M$_{\odot}$ yr$^{-1}$ (Sec. \ref{S:SFRM});
(7) Hubble type based on by-eye morphological classification from \citet{Walcher2014};
(8) Bar class based on by-eye morphological classification from \citet{Walcher2014}, where A--non-barred galaxy, B--barred galaxy and AB -- unsure bar in the galaxy (Sec. \ref{SS:morph});
(9) Quenching stages (QS) of the 238 CALIFA galaxies, where SF -- Star-forming, QnR -- quiescent-nuclear-ring; cQ - centrally-quiescent; MX -- mixed; nR -- nearly-retired and  fR -- fully-retired (Sec. \ref{S:2DELC});
(10) Reliability of the QS classification, where 1  -- sure class , 2 -- unsure class due to complex structure of the galaxy, 3 -- unsure class due to bad data (Sec. \ref{SS:uncertainties}). 
(11) Nuclear activity groups for our sample, where nonA -- non-active galaxy, wAGN--weak AGN and sAGN -- strong AGN  (Sec. \ref{SS:NA_criteria}). 
(12) Reliability of the NA classification, where S: sure class based on the information of the three BPT diagrams and U: unsure class based on the information of two from the three BPT diagrams (Sec. \ref{SS:uncertainties}). }
\label{tab:long} \\
\hline
galaxy  &  D & Re  & $\log$ M$_{*}$ & $\log$ SFR & $\Delta$ SFR & type & bar& QS & flag-QS & NA & flag-NA \\
        & Mpc & arcsec  & M$_{\odot}$ & M$_{\odot}$ yr$^{-1}$  & M$_{\odot}$ yr$^{-1}$  & & & & & &  \\
(1) & (2) & (3) & (4) & (5) & (6) & (7) & (8) & (9) & (10) & (11) & (12)  \\
\hline
\endfirsthead

\multicolumn{12}{c}%
{{\bfseries \tablename\ \thetable{} -- continued from previous page}} \\
\hline 
galaxy  &  D & Re & $\log$ M$_{*}$ & $\log$ SFR & $\Delta$ SFR & type & bar & QS & flag-QS & NA & flag-NA \\
        & Mpc & arcsec & M$_{\odot}$ & M$_{\odot}$ yr$^{-1}$  & M$_{\odot}$ yr$^{-1}$ & & & & & &  \\
(1) & (2) & (3) & (4) & (5) & (6) & (7) & (8) & (9) & (10) & (11) & (12)   \\

\hline 
\endhead

\hline \multicolumn{12}{|r|}{{Continued on next page}} \\ \hline
\endfoot

\hline \hline
\endlastfoot
IC0480 & 61.7 & 14.0 & 9.988 & 0.145 & -0.016 & Sc & AB & SF & 1 & nonA & S \\
IC0540 & 26.1 & 12.1 & 9.736 & -0.903 & -0.869 & Sab & AB & MX & 1 & sAGN & S \\
IC0674 & 103.4 & 9.2 & 10.997 & -0.096 & -1.033 & Sab & B & MX & 2 & nonA & S \\
IC0944 & 95.8 & 13.6 & 11.108 & 0.026 & -0.996 & Sab & A & MX & 1 & nonA & S \\
IC1079 & 119.3 & 18.2 & 11.252 & -0.628 & -1.762 & E4 & A & nR & 2 & nonA & S \\
IC1151 & 31.0 & 19.9 & 9.521 & 0.006 & 0.206 & Scd & B & SF & 1 & nonA & S \\
IC1256 & 67.0 & 10.1 & 10.397 & 0.157 & -0.319 & Sb & AB & SF & 1 & nonA & S \\
IC1528 & 53.3 & 16.0 & 10.134 & 0.428 & 0.155 & Sbc & AB & SF & 1 & nonA & S \\
IC1652 & 72.9 & 8.3 & 10.502 & -1.479 & -2.035 & S0a & A & fR & 1 & nonA & S \\
IC1755 & 109.9 & 16.4 & 10.977 & -0.507 & -1.429 & Sb & A & MX & 1 & nonA & S \\
IC2101 & 60.7 & 15.8 & 10.164 & 0.425 & 0.130 & Scd & AB & SF & 1 & nonA & S \\
IC2247 & 57.7 & 15.2 & 10.429 & 0.154 & -0.345 & Sab & A & SF & 1 & sAGN & S \\
IC2487 & 58.3 & 15.8 & 10.326 & 0.202 & -0.219 & Sc & AB & SF & 1 & nonA & S \\
IC4566 & 81.0 & 12.7 & 10.779 & -0.025 & -0.794 & Sb & B & MX & 1 & nonA & S \\
IC5309 & 59.6 & 15.9 & 10.109 & 0.164 & -0.089 & Sc & AB & SF & 2 & nonA & S \\
IC5376 & 71.5 & 11.1 & 10.584 & -0.254 & -0.873 & Sb & A & cQ & 1 & nonA & S \\
MCG-01-54-016 & 42.1 & 16.8 & 9.156 & -0.528 & -0.048 & Scd & A & SF & 1 & nonA & S \\
MCG-02-02-030 & 48.9 & 12.4 & 10.291 & -0.103 & -0.497 & Sb & AB & QnR & 1 & sAGN & U \\
MCG-02-02-040 & 49.8 & 16.1 & 9.952 & 0.108 & -0.025 & Scd & AB & SF & 1 & nonA & S \\
MCG-02-03-015 & 79.8 & 12.7 & -- & -- & -- & Sab & AB & SF & 3 & nonA & S \\
MCG-02-51-004 & 79.3 & 11.6 & 10.650 & 0.402 & -0.268 & Sb & A & SF & 1 & nonA & S \\
NGC0001 & 64.7 & 11.7 & 10.691 & 0.494 & -0.207 & Sbc & A & SF & 1 & nonA & S \\
NGC0023 & 64.9 & 7.8 & 10.935 & 0.983 & 0.094 & Sb & B & SF & 1 & nonA & S \\
NGC0155 & 85.8 & 9.2 & 10.884 & -1.809 & -2.659 & E1 & A & fR & 1 & nonA & S \\
NGC0160 & 74.0 & 16.8 & 11.047 & 0.050 & -0.926 & Sa & A & cQ & 2 & nonA & S \\
NGC0171 & 53.9 & 18.0 & 10.641 & -0.228 & -0.890 & Sb & B & cQ & 1 & nonA & S \\
NGC0177 & 52.7 & 16.2 & -- & -- & -- & Sab & A & QnR & 2 & nonA & S \\
NGC0192 & 57.9 & 14.2 & 10.750 & 0.348 & -0.399 & Sab & AB & MX & 1 & nonA & S \\
NGC0214 & 64.2 & 14.8 & 10.801 & 0.715 & -0.071 & Sbc & AB & SF & 1 & wAGN & S \\
NGC0216 & 21.4 & 10.9 & 9.185 & -0.404 & 0.054 & Sd & A & SF & 1 & nonA & S \\
NGC0217 & 55.2 & 19.2 & 10.832 & -0.258 & -1.068 & Sa & A & MX & 1 & nonA & S \\
NGC0237 & 58.4 & 10.1 & 10.195 & 0.507 & 0.187 & Sc & B & SF & 1 & nonA & S \\
NGC0257 & 73.7 & 16.2 & 10.842 & 0.941 & 0.124 & Sc & A & SF & 1 & nonA & S \\
NGC0429 & 78.0 & 6.7 & 10.736 & -1.448 & -2.184 & Sa & A & fR & 1 & nonA & S \\
NGC0444 & 68.3 & 15.2 & 9.794 & 0.090 & 0.080 & Scd & A & SF & 1 & nonA & S \\
NGC0499 & 62.3 & 9.2 & 11.240 & -1.140 & -2.265 & E5 & A & fR & 1 & nonA & S \\
NGC0504 & 59.9 & 6.8 & 10.587 & -1.410 & -2.032 & S0 & A & fR & 1 & nonA & S \\
NGC0517 & 59.6 & 5.8 & 10.644 & -1.509 & -2.175 & S0 & A & fR & 1 & nonA & S \\
NGC0528 & 67.8 & 8.2 & 10.932 & -0.970 & -1.857 & S0 & A & nR & 1 & nonA & S \\
NGC0529 & 68.1 & 9.9 & 11.079 & -0.977 & -1.977 & E4 & A & fR & 1 & nonA & S \\
NGC0551 & 73.2 & 15.1 & 10.565 & 0.200 & -0.404 & Sbc & AB & QnR & 1 & nonA & S \\
NGC0681 & 24.3 & 16.6 & 10.160 & -0.245 & -0.537 & Sa & AB & MX & 2 & nonA & S \\
NGC0741 & 77.0 & 15.4 & 11.434 & -0.839 & -2.113 & E1 & A & nR & 1 & nonA & S \\
NGC0755 & 22.6 & 17.8 & 9.300 & -0.298 & 0.072 & Scd & B & SF & 1 & nonA & S \\
NGC0768 & 96.7 & 15.2 & 10.579 & 0.450 & -0.165 & Sc & B & SF & 1 & nonA & S \\
NGC0774 & 64.4 & 7.4 & 10.763 & -0.992 & -1.749 & S0 & A & nR & 1 & nonA & S \\
NGC0776 & 68.8 & 13.2 & 10.803 & 0.382 & -0.406 & Sb & B & SF & 1 & nonA & S \\
NGC0781 & 48.7 & 7.2 & -- & -- & -- & Sa & A & fR & 3 & nonA & S \\
NGC0810 & 106.0 & 10.3 & 11.357 & -0.350 & -1.564 & E5 & A & nR & 2 & nonA & S \\
NGC0932 & 56.9 & 14.0 & 10.957 & -0.454 & -1.360 & S0a & A & nR & 2 & nonA & S \\
NGC1056 & 22.4 & 8.0 & 10.017 & 0.093 & -0.090 & Sa & A & SF & 1 & nonA & S \\
NGC1060 & 72.5 & 13.0 & 11.472 & -0.654 & -1.957 & E3 & A & fR & 1 & nonA & S \\
NGC1167 & 69.1 & 13.4 & 11.231 & -0.320 & -1.437 & S0 & A & nR & 1 & sAGN & U \\
NGC1349 & 90.2 & 11.1 & 11.112 & -0.128 & -1.154 & E6 & A & MX & 2 & nonA & S \\
NGC1542 & 50.3 & 8.9 & 10.314 & 0.054 & -0.357 & Sab & AB & MX & 1 & nonA & S \\
NGC1645 & 65.9 & 12.2 & 10.764 & -0.481 & -1.239 & S0a & B & MX & 1 & nonA & S \\
NGC1677 & 36.5 & 15.2 & 9.467 & -0.195 & 0.046 & Scd & AB & SF & 1 & nonA & S \\
NGC2253 & 50.2 & 14.0 & 10.518 & 0.567 & -0.001 & Sbc & B & SF & 1 & nonA & S \\
NGC2347 & 61.9 & 12.0 & 10.681 & 0.653 & -0.041 & Sbc & AB & SF & 1 & nonA & S \\
NGC2410 & 63.7 & 16.1 & 10.767 & 0.267 & -0.494 & Sb & AB & SF & 1 & sAGN & S \\
NGC2449 & 66.3 & 11.9 & 10.854 & -0.013 & -0.840 & Sab & AB & MX & 1 & nonA & S \\
NGC2476 & 51.0 & 6.9 & 10.673 & -0.771 & -1.458 & E6 & A & fR & 1 & nonA & S \\
NGC2481 & 28.6 & 14.8 & 10.542 & -1.515 & -2.102 & S0 & A & MX & 1 & nonA & S \\
NGC2486 & 62.8 & 19.4 & -- & -- & -- & Sab & B & MX & 1 & nonA & S \\
NGC2553 & 63.5 & 8.6 & 10.591 & -1.038 & -1.662 & Sb & AB & fR & 1 & nonA & S \\
NGC2554 & 56.0 & 15.3 & 11.032 & -0.418 & -1.382 & S0a & A & nR & 1 & wAGN & S \\
NGC2592 & 26.2 & 6.8 & 10.234 & -2.309 & -2.659 & E4 & A & fR & 1 & nonA & S \\
NGC2604 & 27.8 & 18.0 & 9.343 & -0.226 & 0.111 & Sd & B & SF & 1 & nonA & S \\
NGC2639 & 46.1 & 10.9 & 10.853 & -0.191 & -1.018 & Sa & A & MX & 1 & sAGN & S \\
NGC2730 & 51.1 & 18.2 & 9.931 & 0.347 & 0.231 & Scd & B & SF & 1 & nonA & S \\
NGC2880 & 22.4 & 10.0 & 10.324 & -2.292 & -2.710 & E7 & AB & fR & 1 & nonA & S \\
NGC2906 & 27.6 & 12.2 & 10.261 & -0.098 & -0.469 & Sbc & A & cQ & 2 & nonA & S \\
NGC2916 & 50.0 & 18.7 & 10.455 & 0.314 & -0.206 & Sbc & A & cQ & 1 & nonA & S \\
NGC2918 & 93.1 & 7.3 & 10.986 & -1.247 & -2.176 & E6 & A & fR & 1 & nonA & S \\
NGC3057 & 22.9 & 18.5 & 8.775 & -0.806 & -0.033 & Sdm & B & SF & 1 & nonA & S \\
NGC3106 & 84.5 & 11.8 & 11.088 & -0.144 & -1.151 & Sab & A & MX & 2 & nonA & S \\
NGC3160 & 94.7 & 11.0 & 10.706 & -0.508 & -1.220 & Sab & AB & MX & 1 & sAGN & S \\
NGC3300 & 40.9 & 10.1 & 10.484 & -1.560 & -2.102 & S0a & B & fR & 1 & nonA & S \\
NGC3381 & 22.1 & 16.2 & 9.365 & -0.143 & 0.177 & Sd & B & SF & 1 & nonA & S \\
NGC3615 & 90.9 & 6.3 & 11.216 & -0.837 & -1.943 & E5 & A & fR & 1 & nonA & S \\
NGC3811 & 43.2 & 15.3 & 10.304 & 0.202 & -0.201 & Sbc & B & QnR & 1 & nonA & S \\
NGC3815 & 50.4 & 10.0 & 10.406 & 0.649 & 0.167 & Sbc & A & SF & 1 & nonA & S \\
NGC3994 & 42.3 & 5.7 & 10.314 & 0.598 & 0.187 & Sbc & AB & SF & 1 & nonA & S \\
NGC4003 & 88.7 & 8.7 & 10.959 & -0.156 & -1.063 & S0a & B & MX & 2 & nonA & S \\
NGC4047 & 47.5 & 12.9 & 10.506 & 0.620 & 0.060 & Sbc & A & SF & 1 & nonA & S \\
NGC4149 & 43.1 & 11.7 & 10.202 & -0.373 & -0.698 & Sa & AB & MX & 1 & nonA & S \\
NGC4185 & 53.4 & 22.4 & 10.555 & -0.030 & -0.626 & Sbc & AB & cQ & 1 & nonA & S \\
NGC4210 & 39.0 & 19.2 & 10.180 & -0.033 & -0.341 & Sb & B & cQ & 1 & nonA & S \\
NGC4470 & 31.1 & 12.0 & 9.781 & 0.202 & 0.201 & Sc & A & SF & 1 & nonA & S \\
NGC4644 & 69.0 & 13.4 & 10.410 & -0.024 & -0.510 & Sb & A & cQ & 1 & nonA & S \\
NGC4711 & 56.1 & 12.5 & 10.511 & 0.672 & 0.109 & Sbc & A & SF & 1 & nonA & S \\
NGC4816 & 94.9 & 10.3 & 11.067 & -0.605 & -1.596 & E1 & A & nR & 1 & nonA & S \\
NGC4956 & 65.6 & 6.8 & 10.873 & -0.830 & -1.672 & E1 & A & fR & 1 & nonA & S \\
NGC4961 & 35.0 & 11.1 & 9.478 & 0.070 & 0.302 & Scd & B & SF & 1 & nonA & S \\
NGC5000 & 77.1 & 14.5 & 10.667 & 0.233 & -0.450 & Sbc & B & QnR & 2 & nonA & S \\
NGC5029 & 120.9 & 9.5 & 11.423 & -0.474 & -1.740 & E6 & A & nR & 1 & nonA & S \\
NGC5056 & 77.0 & 13.2 & -- & -- & -- & Sc & AB & SF & 1 & nonA & S \\
NGC5218 & 41.9 & 16.1 & 10.445 & 0.083 & -0.429 & Sab & B & MX & 1 & nonA & S \\
NGC5378 & 42.7 & 16.6 & 10.341 & -0.976 & -1.408 & Sb & B & MX & 2 & nonA & S \\
NGC5480 & 27.0 & 15.1 & 9.716 & 0.172 & 0.221 & Scd & A & SF & 1 & nonA & S \\
NGC5485 & 28.1 & 12.8 & 10.530 & -1.780 & -2.358 & E5 & A & fR & 1 & nonA & S \\
NGC5520 & 27.3 & 11.0 & 9.691 & -0.041 & 0.027 & Sbc & A & SF & 1 & nonA & S \\
NGC5614 & 54.4 & 12.3 & 11.109 & -0.098 & -1.122 & Sa & A & nR & 1 & wAGN & S \\
NGC5630 & 37.7 & 13.3 & 9.513 & 0.284 & 0.490 & Sdm & B & SF & 1 & nonA & S \\
NGC5633 & 33.5 & 11.2 & 10.012 & 0.325 & 0.146 & Sbc & A & SF & 1 & nonA & S \\
NGC5657 & 54.5 & 10.7 & 10.260 & 0.190 & -0.179 & Sbc & B & SF & 1 & nonA & S \\
NGC5682 & 32.8 & 25.2 & 9.109 & -0.483 & 0.034 & Scd & B & SF & 1 & nonA & S \\
NGC5720 & 108.4 & 12.4 & 10.947 & 0.120 & -0.779 & Sbc & B & cQ & 1 & nonA & S \\
NGC5732 & 52.7 & 13.2 & 9.726 & 0.026 & 0.067 & Sbc & A & SF & 1 & nonA & S \\
NGC5784 & 75.1 & 8.6 & 11.054 & -0.068 & -1.048 & S0 & A & nR & 1 & nonA & S \\
NGC5876 & 46.9 & 11.6 & 10.654 & -0.833 & -1.506 & S0a & B & nR & 2 & nonA & S \\
NGC5888 & 121.3 & 12.0 & 11.273 & 0.360 & -0.789 & Sb & B & cQ & 2 & nonA & S \\
NGC5908 & 47.3 & 16.2 & 10.941 & -0.051 & -0.945 & Sa & A & MX & 1 & nonA & S \\
NGC5971 & 61.2 & 9.8 & 10.079 & -0.421 & -0.651 & Sb & AB & MX & 2 & wAGN & S \\
NGC5980 & 57.1 & 13.2 & 10.666 & 0.848 & 0.166 & Sbc & A & SF & 1 & nonA & S \\
NGC5987 & 43.5 & 19.7 & 10.753 & -0.796 & -1.545 & Sa & A & nR & 1 & nonA & S \\
NGC6020 & 60.4 & 7.3 & 10.808 & -1.451 & -2.242 & E4 & A & fR & 1 & nonA & S \\
NGC6021 & 66.2 & 5.9 & 10.896 & -1.282 & -2.142 & E5 & A & fR & 1 & nonA & S \\
NGC6032 & 60.1 & 14.8 & 10.281 & -0.297 & -0.683 & Sbc & B & MX & 2 & nonA & S \\
NGC6060 & 62.3 & 22.3 & 10.795 & 0.676 & -0.106 & Sb & A & SF & 1 & nonA & S \\
NGC6063 & 40.0 & 17.9 & 9.924 & -0.372 & -0.483 & Sbc & A & SF & 1 & nonA & S \\
NGC6081 & 72.0 & 10.0 & 10.935 & -0.546 & -1.435 & S0a & A & nR & 1 & sAGN & U \\
NGC6125 & 68.6 & 8.0 & 11.202 & -1.426 & -2.521 & E1 & A & fR & 1 & nonA & S \\
NGC6132 & 69.2 & 14.2 & 10.099 & 0.325 & 0.080 & Sbc & A & SF & 1 & nonA & S \\
NGC6146 & 123.0 & 7.9 & 11.463 & -0.767 & -2.063 & E5 & A & nR & 1 & wAGN & U \\
NGC6150 & 121.6 & 7.2 & 11.252 & -0.744 & -1.878 & E7 & A & fR & 1 & nonA & S \\
NGC6168 & 36.1 & 18.1 & 9.596 & -0.031 & 0.111 & Sc & AB & SF & 1 & nonA & S \\
NGC6173 & 122.5 & 13.9 & 11.681 & -0.576 & -2.040 & E6 & A & nR & 2 & nonA & S \\
NGC6186 & 41.9 & 11.9 & 10.454 & 0.183 & -0.336 & Sb & B & MX & 1 & nonA & S \\
NGC6278 & 40.7 & 8.4 & 10.676 & -1.970 & -2.660 & S0a & AB & fR & 1 & nonA & S \\
NGC6301 & 116.8 & 21.5 & 10.870 & 0.524 & -0.315 & Sbc & A & SF & 1 & nonA & S \\
NGC6310 & 49.5 & 15.3 & 10.451 & -0.353 & -0.870 & Sb & A & cQ & 1 & nonA & S \\
NGC6314 & 92.9 & 9.4 & 10.921 & -0.195 & -1.073 & Sab & A & nR & 1 & nonA & S \\
NGC6478 & 95.6 & 15.5 & 10.986 & 0.796 & -0.133 & Sc & A & SF & 1 & nonA & S \\
NGC6497 & 87.2 & 12.3 & 10.885 & -0.050 & -0.901 & Sab & B & cQ & 1 & nonA & S \\
NGC6515 & 96.7 & 10.8 & 11.086 & -0.945 & -1.951 & E3 & A & nR & 1 & nonA & S \\
NGC6762 & 43.0 & 10.3 & 10.205 & -1.714 & -2.042 & Sab & A & nR & 1 & wAGN & S \\
NGC6941 & 87.2 & 16.6 & 10.982 & 0.072 & -0.853 & Sb & B & QnR & 1 & nonA & S \\
NGC6945 & 53.9 & 8.8 & 9.385 & -2.882 & -2.578 & S0 & B & fR & 1 & nonA & S \\
NGC6978 & 84.5 & 14.9 & 10.814 & 0.198 & -0.598 & Sb & AB & MX & 2 & nonA & S \\
NGC7025 & 70.8 & 13.0 & 11.242 & -0.629 & -1.754 & S0a & A & nR & 1 & nonA & S \\
NGC7047 & 81.7 & 17.3 & 10.721 & 0.324 & -0.401 & Sbc & B & cQ & 1 & nonA & S \\
NGC7194 & 112.7 & 3.6 & 11.345 & -1.158 & -2.363 & E3 & A & fR & 1 & nonA & S \\
NGC7311 & 64.3 & 8.6 & 11.117 & 0.105 & -0.925 & Sa & A & cQ & 1 & nonA & S \\
NGC7321 & 100.5 & 13.0 & 11.028 & 0.495 & -0.466 & Sbc & B & QnR & 1 & wAGN & U \\
NGC7364 & 68.5 & 8.0 & 10.706 & 0.483 & -0.230 & Sab & A & SF & 1 & nonA & S \\
NGC7466 & 105.5 & 12.7 & 10.709 & 0.455 & -0.260 & Sbc & A & SF & 1 & sAGN & S \\
NGC7489 & 88.0 & 16.2 & 10.493 & 0.748 & 0.199 & Sbc & A & SF & 1 & nonA & S \\
NGC7549 & 67.3 & 19.1 & 10.510 & 0.780 & 0.218 & Sbc & B & SF & 2 & nonA & S \\
NGC7550 & 71.9 & 11.2 & 11.157 & -1.067 & -2.128 & E4 & A & nR & 1 & wAGN & U \\
NGC7562 & 51.4 & 10.5 & 11.187 & -0.979 & -2.063 & E4 & A & fR & 1 & nonA & S \\
NGC7563 & 59.4 & 7.5 & 11.026 & -1.391 & -2.350 & Sa & B & fR & 1 & nonA & S \\
NGC7591 & 69.9 & 11.7 & 10.731 & 0.703 & -0.030 & Sbc & B & SF & 1 & nonA & S \\
NGC7608 & 50.1 & 12.4 & 9.946 & 0.043 & -0.085 & Sbc & A & SF & 1 & nonA & S \\
NGC7611 & 46.6 & 9.0 & 10.652 & -0.999 & -1.671 & S0 & A & fR & 1 & nonA & S \\
NGC7619 & 53.6 & 11.2 & 11.267 & -1.075 & -2.220 & E3 & A & fR & 1 & nonA & S \\
NGC7623 & 53.3 & 6.8 & 10.732 & -1.353 & -2.087 & S0 & A & fR & 1 & nonA & S \\
NGC7625 & 24.7 & 10.7 & 9.963 & 0.210 & 0.069 & Sa & A & SF & 2 & nonA & S \\
NGC7631 & 53.5 & 13.8 & 10.507 & 0.009 & -0.551 & Sb & A & QnR & 1 & nonA & S \\
NGC7653 & 60.7 & 11.8 & 10.512 & 0.471 & -0.093 & Sb & A & cQ & 1 & nonA & S \\
NGC7671 & 58.7 & 6.4 & 10.798 & -2.010 & -2.794 & S0 & A & fR & 1 & nonA & S \\
NGC7683 & 53.2 & 9.6 & 10.972 & -1.129 & -2.048 & S0 & A & fR & 1 & nonA & S \\
NGC7711 & 57.8 & 12.0 & 10.835 & -1.047 & -1.860 & E7 & A & MX & 2 & nonA & S \\
NGC7716 & 37.0 & 12.3 & 10.410 & -0.231 & -0.716 & Sb & A & cQ & 1 & nonA & S \\
NGC7722 & 57.4 & 12.9 & 10.934 & 0.085 & -0.804 & Sab & A & MX & 2 & nonA & S \\
NGC7738 & 94.2 & 13.1 & 11.031 & 0.716 & -0.247 & Sb & B & MX & 1 & nonA & S \\
NGC7787 & 92.8 & 11.2 & 10.573 & 0.215 & -0.396 & Sab & AB & QnR & 1 & nonA & S \\
NGC7819 & 70.4 & 12.4 & 10.068 & 0.326 & 0.104 & Sc & A & SF & 1 & nonA & S \\
NGC7824 & 85.7 & 9.6 & 11.045 & -0.224 & -1.198 & Sab & A & nR & 1 & sAGN & U \\
UGC00005 & 101.0 & 13.0 & 10.843 & 0.795 & -0.023 & Sbc & A & SF & 1 & sAGN & U \\
UGC00029 & 122.7 & 10.4 & -- & -- & -- & E1 & A & nR & 1 & nonA & S \\
UGC00036 & 87.9 & 11.2 & 10.897 & -0.083 & -0.943 & Sab & AB & MX & 1 & nonA & S \\
UGC00148 & 59.7 & 4.5 & 10.148 & 0.831 & 0.547 & Sc & A & SF & 1 & nonA & S \\
UGC00312 & 61.4 & 15.7 & 9.851 & 0.439 & 0.384 & Sd & B & SF & 1 & nonA & S \\
UGC00809 & 59.6 & 11.7 & 9.586 & -0.107 & 0.042 & Scd & A & SF & 1 & nonA & S \\
UGC00987 & 65.8 & 9.8 & 10.724 & -0.303 & -1.030 & Sa & AB & SF & 1 & sAGN & S \\
UGC01057 & 88.7 & 10.8 & 10.258 & 0.328 & -0.040 & Sc & AB & SF & 1 & nonA & S \\
UGC01271 & 70.8 & 6.1 & 10.772 & -0.724 & -1.487 & S0a & B & fR & 1 & nonA & S \\
UGC02222 & 70.6 & 10.3 & 10.696 & -1.237 & -1.942 & S0a & AB & fR & 1 & nonA & S \\
UGC02229 & 100.5 & 16.3 & 10.918 & -0.544 & -1.420 & S0a & A & nR & 1 & nonA & S \\
UGC02403 & 56.8 & 12.3 & 10.395 & 0.153 & -0.320 & Sb & B & SF & 1 & nonA & S \\
UGC03253 & 58.8 & 19.2 & 10.360 & -0.181 & -0.628 & Sb & B & QnR & 1 & nonA & S \\
UGC03539 & 46.7 & 19.2 & 9.654 & -0.072 & 0.025 & Sc & AB & SF & 1 & nonA & S \\
UGC03995 & 64.4 & 18.0 & 10.933 & -0.003 & -0.891 & Sb & B & QnR & 2 & sAGN & S \\
UGC04029 & 60.1 & 17.9 & 10.250 & 0.148 & -0.214 & Sc & AB & SF & 1 & nonA & S \\
UGC04132 & 71.0 & 14.3 & 10.748 & 0.893 & 0.147 & Sbc & AB & SF & 1 & nonA & S \\
UGC04145 & 62.0 & 8.6 & 10.468 & 0.509 & -0.021 & Sa & AB & MX & 1 & sAGN & S \\
UGC04197 & 61.1 & 15.6 & 10.497 & -0.342 & -0.894 & Sab & AB & cQ & 1 & nonA & S \\
UGC04280 & 48.8 & 9.9 & 10.190 & -0.318 & -0.634 & Sb & AB & SF & 1 & nonA & S \\
UGC04308 & 47.8 & 20.6 & 10.021 & 0.191 & 0.006 & Sc & B & SF & 1 & nonA & S \\
UGC05108 & 110.2 & 12.9 & 10.927 & 0.164 & -0.719 & Sb & B & QnR & 1 & nonA & S \\
UGC05113 & 94.7 & 8.1 & 10.913 & -1.213 & -2.085 & S0a & AB & nR & 1 & nonA & S \\
UGC05598 & 76.2 & 11.9 & 10.066 & 0.219 & -0.001 & Sb & A & SF & 1 & nonA & S \\
UGC05771 & 101.6 & 6.9 & 11.141 & -0.318 & -1.366 & E6 & A & nR & 1 & nonA & S \\
UGC05990 & 21.3 & 7.7 & -- & -- & -- & Sc & A & SF & 1 & nonA & S \\
UGC06036 & 89.0 & 10.2 & 11.039 & -0.324 & -1.293 & Sa & A & MX & 1 & nonA & S \\
UGC06312 & 85.2 & 11.5 & 10.793 & -0.161 & -0.941 & Sab & A & MX & 1 & nonA & S \\
UGC07012 & 42.1 & 10.4 & 9.266 & -0.175 & 0.220 & Scd & AB & SF & 1 & nonA & S \\
UGC07145 & 91.3 & 12.4 & 10.343 & 0.059 & -0.374 & Sbc & A & SF & 1 & nonA & S \\
UGC08107 & 115.0 & 15.4 & 11.059 & 0.834 & -0.151 & Sa & A & SF & 1 & nonA & S \\
UGC08231 & 35.2 & 13.9 & 9.035 & -0.216 & 0.358 & Sd & AB & SF & 1 & nonA & S \\
UGC08234 & 112.6 & 4.7 & 11.040 & 0.070 & -0.900 & S0 & A & fR & 1 & nonA & S \\
UGC08733 & 33.2 & 19.8 & 9.072 & -0.788 & -0.243 & Sdm & B & SF & 1 & nonA & S \\
UGC08778 & 45.8 & 9.8 & 9.994 & -0.594 & -0.759 & Sb & A & MX & 1 & nonA & S \\
UGC08781 & 104.3 & 10.9 & 10.978 & -0.138 & -1.060 & Sb & B & cQ & 1 & nonA & S \\
UGC09476 & 46.2 & 17.2 & 10.032 & 0.195 & 0.001 & Sbc & A & SF & 1 & nonA & S \\
UGC09537 & 122.1 & 16.5 & 11.054 & 0.212 & -0.770 & Sb & A & cQ & 1 & nonA & S \\
UGC09542 & 76.4 & 15.0 & 10.185 & 0.227 & -0.085 & Sc & A & SF & 1 & nonA & S \\
UGC09665 & 36.8 & 14.5 & 9.798 & 0.069 & 0.055 & Sb & A & SF & 1 & nonA & S \\
UGC09873 & 78.8 & 16.3 & 9.905 & -0.076 & -0.172 & Sb & A & SF & 1 & nonA & S \\
UGC09892 & 79.5 & 13.0 & 10.137 & 0.165 & -0.110 & Sbc & A & SF & 1 & nonA & S \\
UGC10097 & 83.8 & 7.8 & 11.242 & -0.973 & -2.099 & E5 & A & nR & 1 & nonA & S \\
UGC10123 & 53.8 & 10.6 & 10.306 & 0.328 & -0.077 & Sab & A & SF & 1 & nonA & S \\
UGC10205 & 91.6 & 14.6 & 10.952 & 0.353 & -0.550 & S0a & A & nR & 1 & nonA & S \\
UGC10257 & 54.5 & 20.1 & 10.511 & 0.858 & 0.294 & Sbc & A & SF & 1 & nonA & S \\
UGC10331 & 63.7 & 17.3 & 9.840 & 0.184 & 0.138 & Sc & AB & SF & 1 & nonA & S \\
UGC10337 & 121.1 & 11.3 & 10.967 & 0.540 & -0.375 & Sb & A & cQ & 1 & nonA & S \\
UGC10384 & 69.3 & 8.6 & 10.234 & 0.639 & 0.290 & Sb & A & SF & 1 & nonA & S \\
UGC10388 & 64.9 & 6.7 & 10.473 & -0.718 & -1.252 & Sa & AB & MX & 2 & nonA & S \\
UGC10650 & 42.5 & 11.3 & 9.170 & -0.218 & 0.252 & Scd & A & SF & 1 & nonA & S \\
UGC10693 & 116.9 & 10.2 & 11.272 & -0.655 & -1.804 & E7 & AB & fR & 1 & nonA & S \\
UGC10695 & 116.5 & 11.7 & 11.190 & -0.733 & -1.819 & E5 & A & nR & 1 & nonA & S \\
UGC10710 & 117.3 & 13.9 & 10.885 & 1.275 & 0.424 & Sb & A & SF & 1 & nonA & S \\
UGC10796 & 44.8 & 11.4 & 9.199 & -0.259 & 0.188 & Scd & AB & SF & 1 & nonA & S \\
UGC10811 & 122.5 & 10.7 & 10.955 & 0.468 & -0.437 & Sb & B & cQ & 1 & nonA & S \\
UGC10905 & 109.7 & 17.6 & 11.277 & -0.342 & -1.495 & S0a & A & nR & 1 & nonA & S \\
UGC10972 & 66.1 & 15.0 & 10.329 & -0.219 & -0.642 & Sbc & A & cQ & 1 & nonA & S \\
UGC11228 & 81.9 & 6.2 & 10.962 & -1.303 & -2.214 & S0 & B & nR & 1 & nonA & S \\
UGC11717 & 89.1 & 11.6 & 10.824 & 0.802 & -0.001 & Sab & A & MX & 1 & sAGN & U \\
UGC12054 & 30.6 & 10.0 & 8.993 & -0.557 & 0.048 & Sc & A & SF & 1 & nonA & S \\
UGC12127 & 116.2 & 16.5 & 11.339 & -0.702 & -1.903 & E1 & A & nR & 1 & nonA & S \\
UGC12185 & 93.9 & 16.5 & 10.693 & -0.128 & -0.831 & Sb & B & cQ & 1 & nonA & S \\
UGC12274 & 107.7 & 20.6 & 11.072 & -0.340 & -1.336 & Sa & A & nR & 1 & nonA & S \\
UGC12308 & 32.9 & 20.3 & 8.916 & -0.571 & 0.094 & Scd & A & SF & 1 & nonA & S \\
UGC12518 & 40.2 & 11.8 & 10.374 & -0.490 & -0.948 & Sb & A & MX & 2 & nonA & S \\
UGC12519 & 62.3 & 12.5 & 10.124 & 0.325 & 0.060 & Sc & AB & SF & 1 & nonA & S \\
UGC12723 & 76.9 & 5.8 & 9.628 & -0.144 & -0.027 & Sc & A & SF & 1 & nonA & S \\
UGC12857 & 35.5 & 13.8 & 9.578 & -0.213 & -0.058 & Sbc & A & SF & 1 & nonA & S \\
\end{longtable}
\end{center}


\begin{table}
\caption{Median values of galaxy properties from the Box-and-Whisker diagrams in Fig. \ref{fig:Box} (see Sec. \ref{S:data} and \ref{S:ELCprop}).
Columns list: 
(1) Quenching stage (QS) of non-active (nonA) galaxies,  where: SF -- Star-forming, QnR -- quiescent-nuclear-ring; cQ - centrally-quiescent; MX -- mixed; nR -- nearly-retired and  fR -- fully-retired (Sec. \ref{S:2DELC});
(2) The median bulge-to-disk ratios for a given EL class;
(3) The median value of the total $r-$band luminosity for a given EL class; 
(4) The median values of the effective radius for a given EL class;
(5) The median stellar mass surface density for a given EL class;
(6) The median total stellar mass of the galaxies for a given EL class;
(7) The median value of the galaxy age for a given EL class;
(8) The median value of galaxy metallicity for a given EL class;
(9) The median value of star formation rate for a given EL class;
(10) The median value of the specific angular momentum within one effective radius for a given EL class; 
(11) The median value of the maximum circular velocity for a given EL class; 
(12) The median value of the total dynamical mass for a given EL class; 
(13) The median value of mass discrepancy factor, f$_{d}=1-($M$^{\mathrm{tot}}_{\mathrm{*}}/$M$^{\mathrm{tot}}_{\mathrm{dyn}})$, for a given EL class;
see text in Sec. \ref{S:data} and \ref{S:ELCprop} for more details.
}
\label{tab:Box-medians} 
\begin{tabular}{|l|l|l|l|l|l|l|l|l|l|l|l|l|}
\hline
QS & B/D & log L$^\mathrm{tot}_\mathrm{r}$ & R$_{\mathrm{e}}$ & log $\mu_{*}$  & log M$^{\mathrm{tot}}_{\mathrm{*}}$  & log $\tau$ & log Z & log SFR & $\lambda_{\mathrm{R_e}}$ & V$_{\mathrm{c,max}}$  & log M$^{\mathrm{tot}}_{\mathrm{dyn}}$  & f$_{d}$ \\
&  &  [L$_{\odot}$] &  [kpc] &  [M$_{\odot}/pc^2$] &  [M$_{\odot}$] &  [yr] &  [Z$_{\odot}$] & [M$_{\odot}\,$ yr$^{-1}$] &  &  [km s$^{-1}$] & & \\
(1) & (2) & (3) & (4) & (5) & (6) & (7) & (8) & (9) & (10) & (11) & (12) & (13) \\
\hline
SF-nonA & 0.08 & 10.03 & 3.41 & 2.38 & 10.07 & 8.80 & -0.24 & 0.20 & 0.60 & 171.85 & 10.77 & 0.80 \\
QnR-nonA & 0.09 & 10.26 & 5.35 & 2.67 & 10.57 & 9.45 & 0.04 & 0.18 & 0.70 & 205.83 & 10.95 & 0.64 \\
cQ-nonA & 0.11 & 10.51 & 4.81 & 2.83 & 10.64 & 9.52 & 0.12 & -0.03 & 0.66 & 237.26 & 11.18 & 0.65 \\
MX-nonA & 0.30 & 10.50 & 3.89 & 3.01 & 10.80 & 9.65 & 0.08 & -0.15 & 0.58 & 245.28 & 11.25 & 0.66 \\
nR-nonA & 0.58 & 10.79 & 4.89 & 3.19 & 11.07 & 9.81 & 0.19 & -0.63 & 0.29 & 343.61 & 11.62 & 0.68 \\
fR-nonA & 0.69 & 10.47 & 2.23 & 3.39 & 10.81 & 9.77 & 0.21 & -1.25 & 0.39 & 326.74 & 11.22 & 0.64 \\
\hline
\end{tabular}
\end{table}


\label{lastpage}
\end{document}